\documentclass[11pt]{article}
\usepackage{amsmath}
\usepackage{amsfonts}
\usepackage{amssymb}
\usepackage{graphicx}
\usepackage{geometry}
\usepackage{hyperref}
\usepackage{multirow}
\usepackage{xcolor}
\usepackage[english]{babel}
\usepackage[bottom]{footmisc}
\usepackage[utf8]{inputenc}
\usepackage[T1]{fontenc}
\title{Design of Experiment in Complex Systems based on Computational Taxonomy.}
\author{Eric Goldman\footnotemark[1] and Fushing Hsieh\footnotemark[1]\footnotemark[2]}
\date{today}
\begin{document}
\maketitle
\footnotetext[1]{Department of Statistics, University of California at Davis, CA, 95616. USA.}
\footnotetext[1]{Correspondences: Fushing H. Department of Statistics, University of California at Davis, email:fhsieh@ucdavis.edu}

\begin{abstract}
Via Computational Taxonomy (CT), we develop Design of Experiment(DoE) based on rigorously redefined constituting ingredients of complex system dynamics: randomness, nonlinearity and even class, through a data-driven constructed Taxonomic Hierarchy. As an opposite quest of Classification without man-made assumptions and structures, we illustrate this new perspective of DoE through a civil engineering complex system: Concrete Compressive Strength (CCS). CT begins by building a Taxonomic Hierarchy as a heterogeneity-vs-homogeneity map framed with a tree geometry to represent CCS-system dynamics. At each internal node of this hierarchy, a heatmap is computed via Scientific Data Analysis (SDA) to reveal locality-embraced heterogeneity through block structured covariate homogeneity annotated with response's locality-split. Only arriving at each ending-node of this hierarchy, coherence of homogeneity is achieved on both response and covariate sides. As such a class of finite sample nature is computationally recognized and confirmed. In contrast, nonlinearity is evidently observed as incoherence of response-vs-covariate homogeneity when comparing two classes located on two distinct branches. This hierarchy explicitly maps out system's randomness and nonlinearity to serve as a scientific basis for any DoE quest. Design of Experiment (DoE) for any designated class is redefined as a search for a covariate subspace that embraces the class-representative randomness and at the same time avoids potential nonlinearities with respect to the rest of classes. This is a brand-new theme of DoE.
\end{abstract}

\section{Introduction}
A study of any complex system \cite{mitchell} ideally must reveal where its component dynamics are and how they collectively operate \cite{anderson}. From a data analysis perspective, this scientific goal can be effectively translated into building a representative heterogeneity-vs-homogeneity map. Each locality of homogeneity stands for a local component of system dynamics, while the composition of locality-specific homogeneity across all localities becomes a map of heterogeneity. Only based on such a heterogeneity-vs-homogeneity map can the primary goal of studying a complex system via data analysis be coherently explained. As such, ``the principle of science'' advocated by quantum computing physicist David Deutsch \cite{deutsch11} can be achieved.

Computational Taxonomy (CT) was recently developed in accord with this principle of science on two small systems: Iris and penguin data sets, with illustrative clarity and precision \cite{CTonIris,CTonpenguin}. In both studies, we constructed heterogeneity-vs-homogeneity maps to reveal two biological complex systems for better understanding and brand-new discoveries. We also demonstrated that the classification task within each system can be resolved as a by-product of this map. Can Design of Experiment (DoE) within a complex system be resolved effectively as well?

Here, DoE is taken as the opposite task of classification. Its goal is simply stated as: {\bf How can we find a subspace of covariate features that will lead to a designated class of the response variable with reliability?} In this paper, we not only provide a positive answer to this question, but also illustrate an entirely new perspective of experimental design within any complex system. The developments throughout this paper are illustrated by reanalyzing a Concrete Compressive Strength (CCS) data set, which plays the role of a pilot study, with one 1D CCS as the response variable and eight 1D covariate quantitative features. This CCS data set, available in the UCI Data Repository, comes from a highly nonlinear civil engineering system \cite{yeh}.

Thus, the DoE question for the CCS system is: How can we find an 8-dimensional range of covariate features that will lead to a designated class of CCS? At present, it seems very difficult, if not impossible, to resolve this DoE question without any man-made modeling assumptions and structures.

The intuitive ideas behind our CT-based DoE within this CCS complex system are narrated as follows. To begin our CT paradigm, a complex system is to be represented by a data-driven structural network of heterogeneity that coherently links and properly positions a collection of locality-based pure randomness, or homogeneity. This heterogeneity-vs-homogeneity map is motivated by Nobel physicist P. W. Anderson's paper \cite{anderson}. Specifically, this map equipped with a tree geometry is called a Taxonomic Hierarchy.

\subsection{Review of SDA for Taxonomic Hierarchy construction.}
A Taxonomic Hierarchy is constructed through Scientific Data Analysis (SDA) computations, which have been rigorously developed and clearly illustrated in \cite{omotayo,CTonIris,CTonpenguin}. Here, only a concise version through a series of data-driven steps is given. First, a Hierarchical Clustering (HC) algorithm is applied on CCS measurements to build a binary HC-tree capturing its distributional patterns and randomness. Such distributional information can be seen by cutting the HC-tree at a tree-level to produce a histogram \cite{FR2018}. This histogram's piecewise-linear distribution approximates CCS's Empirical Cumulative Distribution Function (ECDF). Since the ECDF is supposed to embrace the entire distributional information contained in the collection of CCS measurements, this histogram captures nearly the same distributional information. Further, given the known uniform randomness within all bins, this histogram is taken as an informative categorization scheme for the CCS response variable.

As such, the cut HC-tree becomes the starting tree geometry for constructing the Taxonomic Hierarchy. This construction task relies entirely on SDA. Starting from the top tree-level of this cut HC-tree, a split at its internal node results in two branches: L1 and R1, on its left (L) and right (R) hand sides. That is, the entire data set is split into two parts by defining a binary response variable with two categories: $\{L1, R1\}$. A response(Re)-covariate(Co) dynamic is thereby defined at this internal node, denoted $\circ$. Upon this $Re$-$Co[\circ]$ dynamic, the associative relationships between this binary response variable and the eight 1D covariate features are explored by applying the data-driven exploratory computations of SDA.

Here very briefly, SDA is primarily performed on contingency table of any covariate feature-set against the binary response-variable within the $Re$-$Co[\circ]$ dynamic. To build such a contingency table with two response-category-specific columns, any involved covariate feature-set of any order $k$ with $k=1,..,8$ is categorized by using the HC-tree of its $k$ members. Any row-vector of this contingency table is a candidate piece of associative information. Its finite sample precision is determined by observed and null randomness of this contingency table. See details in \cite{omotayo,CTonIris,CTonpenguin}.

All computed and confirmed pieces of associative information are collected and displayed on a heatmap platform with a fixed format: all study-subjects are arranged along the column-axis, and each row is a binary vector recording the presence or absence of each study-subject that defines a piece of associative information. This heatmap becomes a bipartite network that collectively embraces potentially critical interacting relations between study-subjects and pieces of associative information, as explored below.

By properly permuting the rows and columns in an iterative fashion via the HC algorithm \cite{CF12,FC14}, this matrix lattice of the heatmap comes to be sustained by horizontal and vertical block-chain structures. Each block is framed by one cluster of study-subjects and one cluster of associative information pieces. A horizontal block-chain reveals a specific version of so-called mechanistic dependence among members of its defining associative-information cluster, by sharing various clusters of study-subjects' common presences or absences. This block-chain signals an aspect of system dynamics under the $Re$-$Co[\circ]$ dynamic.

In contrast, a vertical block-chain characterizes the corresponding study-subject cluster through a series of associative-information clusters. In this fashion, each individual is characterized by its participation status across a series of mechanistic dependencies under the $Re$-$Co[\circ]$ dynamic. Importantly, the collective of such vertical block-chains brings out a version of the heterogeneity-vs-homogeneity map pertaining to the $Re$-$Co[\circ]$ dynamic.

If a block-structure is evidently present on this SDA-based heatmap at the top internal node, we confirm the split of L1-vs-R1. In this fashion, we discover CCS-data's heterogeneity of global scales. Otherwise, we declare that the eight covariate features offer no associative relational evidence. After confirming the L1-vs-R1 separation, we go on to explore heterogeneity on the next scale. This task is carried out by descending to the internal node where L1 is separated into L1L2-vs-L1R2, and to the internal node where R1 is separated into R1L2-vs-R1R2.

At L1, the $Re$-$Co[L1]$ dynamic is defined. Upon this dynamic, we likewise perform SDA computations to check whether L1L2 should be separated from L1R2, based on a heatmap that reveals a version of the heterogeneity-vs-homogeneity map under the $Re$-$Co[L1]$ dynamic. If the split is confirmed, we have identified one part of the heterogeneity in $Re$-$Co[\circ]$. We then descend further to the internal nodes of L1L2 and L1R2 at the next tree-level. SDA is likewise applied to the $Re$-$Co[L1L2]$ and $Re$-$Co[L1R2]$ dynamics, respectively.

At R1, we likewise descend to the internal node of R1, where a split, R1L2-vs-R1R2, is seen, and a $Re$-$Co[R1]$ dynamic is defined. We perform SDA computations to check whether R1L2 should be separated from R1R2 based on a heatmap that reveals a version of the heterogeneity-vs-homogeneity map under $Re$-$Co[R1]$. If the split is confirmed, we have identified another part of the heterogeneity in $Re$-$Co[\circ]$, and proceed to the next tree-level. SDA is likewise applied to the $Re$-$Co[R1L2]$ and $Re$-$Co[R1R2]$ dynamics, respectively.

Consequently, any confirmed split at any internal node of any tree-level indicates that heterogeneity has been discovered within the branch housing that internal node, and is revealed through a heatmap pertaining to a locality-specific Re-Co dynamic. Hence, by further exploring potential heterogeneity within sub-branches until such confirmation is denied, we discover all potential multiscale heterogeneity across the internal nodes of the CCS HC-tree.

Most importantly, whenever no split is confirmed at an internal node -- for instance, when few or no associative information pieces are confirmed in SDA computations -- this is taken as a declaration of discovering locality-specific homogeneity. This splitting-ending branch is then taken as a class. Precisely, an identified class means that CCS values within it are essentially non-differentiable with respect to the information provided by all covariate features within the data subset defining that class. Such coherence between response and covariate homogeneity at a locality is the essence of locality-specific randomness underlying and defining a class. {\bf This definition of finite sample nature is a critically new concept in data analysis.}

In summary, through the tree-structured dynamics $Re$-$Co[\circ]$, $Re$-$Co[L1]$, $Re$-$Co[R1]$, $Re$-$Co[L1L2]$, and so on, SDA computing builds tree-structured, internal-node-specific heatmaps that bring out the system's heterogeneity-vs-homogeneity map. This is a tree-structured representation of the CCS system dynamics embedded within the original CCS data set. In this precise sense, the computed Taxonomic Hierarchy maps out the multiscale heterogeneity-vs-homogeneity underlying the complex system.

\subsection{DoE for science.}
Next, we explain the scientific validity of this brand-new experimental design. In particular, we emphasize that the targeted complex system's effective and informative representation -- a Taxonomic Hierarchy coupled with a tree-structured spectrum of heatmaps -- can afford Scientific Experimental Design without relying on any modeling assumptions or structures.

How does this model-free DoE bring out the range of 8-dimensional covariate features? Conceptually, upon each SDA-based heatmap constructed at each internal node at any tree-level, its horizontal block-chains play a key role in constraining regions within the space of the 8 covariate features for CT-based DoE, while its vertical block-chains serve as a basis for selecting targeted study-subjects to fulfill the goal of DoE. The computational protocol for carrying out DoE is detailed and demonstrated in the DoE section.

To close this section, we briefly contrast classic DoE with our CT-based DoE. The simplest form of experimental design in statistics is essentially a model-based optimization \cite{montgomery,box}, in which the system under study is represented by an assumed model equipped with assumptions and structures of homogeneity. The goal is primarily defined and achieved by finding an optimal solution. This top-down modeling approach is, at best, a form of mathematics. It is unscientific when facing real-world systems, because it cannot pass the test of experience \cite{tukey}.

To support this statement, we quote Wikipedia's description of Design of Experiment: \\
``...In general, the design of experiments involves decisions about which aspects of the system to change and which to control, based on hypotheses about the sources of variance in the aspects of the system considered by the experimenter.''\\

First, ``the source of variance'' is meaningful only when an unimodal distributional assumption, such as normality, holds on the global scale. Such a strict, overall version of homogeneity hardly exists in any real-world system, since most systems of interest are complex in nature. Such complexity likely involves categorical response and covariate features, with which variance is not a valid concept.

Second, ``...which aspects of the system to change and which to control...'' is a meaningful act only when the system under study is already well understood. The underlying dynamics of complex systems are sustained by interacting effects of unknown order and unknown form. Without prior knowledge of interacting effects across all covariate features, there is generally no way of knowing which covariate features to change, or how, in most complex systems under study.

Given a completely unknown complex system's dynamics, how can experimental design be carried out scientifically based on data from a pilot study? In this paper, we rigorously address this chief question, which bears on nearly all branches of science. In simple terms, scientific experimental design is defined as:\\
{\bf Figuring out ranges of covariate feature-sets that can collectively lead to a designated response category with high enough reliability.}\\
This version of experimental design is a discovery-oriented task upon the Taxonomic Hierarchy, coupled with tree-structured heterogeneity-vs-homogeneity maps representing the targeted system's dynamics. This is the brand-new DoE based on Computational Taxonomy (CT) applied to pilot data.

We illustrate CT-based Experimental Design via a real-world data set \cite{yeh}: Concrete Compressive Strength (CCS). This data set is available in the UCI Data Repository. A quote from the abstract of the original reference: ``High-performance concrete is a highly complex material, which makes modeling its behavior a very difficult task.'' The author built and employed artificial neural networks (ANN) on the data set for the predictive task. Nonetheless, an ANN is essentially a black box, with no explanations provided alongside its predictions.

As quantum computing physicist David Deutsch put it \cite{deutsch11}: ``Prediction is not, and cannot be, the purpose of science.'' He advocated the principle of science via ``explanation.'' Design of experiment is carried out by scientists on a daily basis, and it has to be scientific. That is why we emphasize scientific validity upon this brand-new DoE proposed in this paper.

\section{Computational Taxonomy}
\begin{figure}[ht!]
 \centering
 \includegraphics[width=1.0\textwidth]{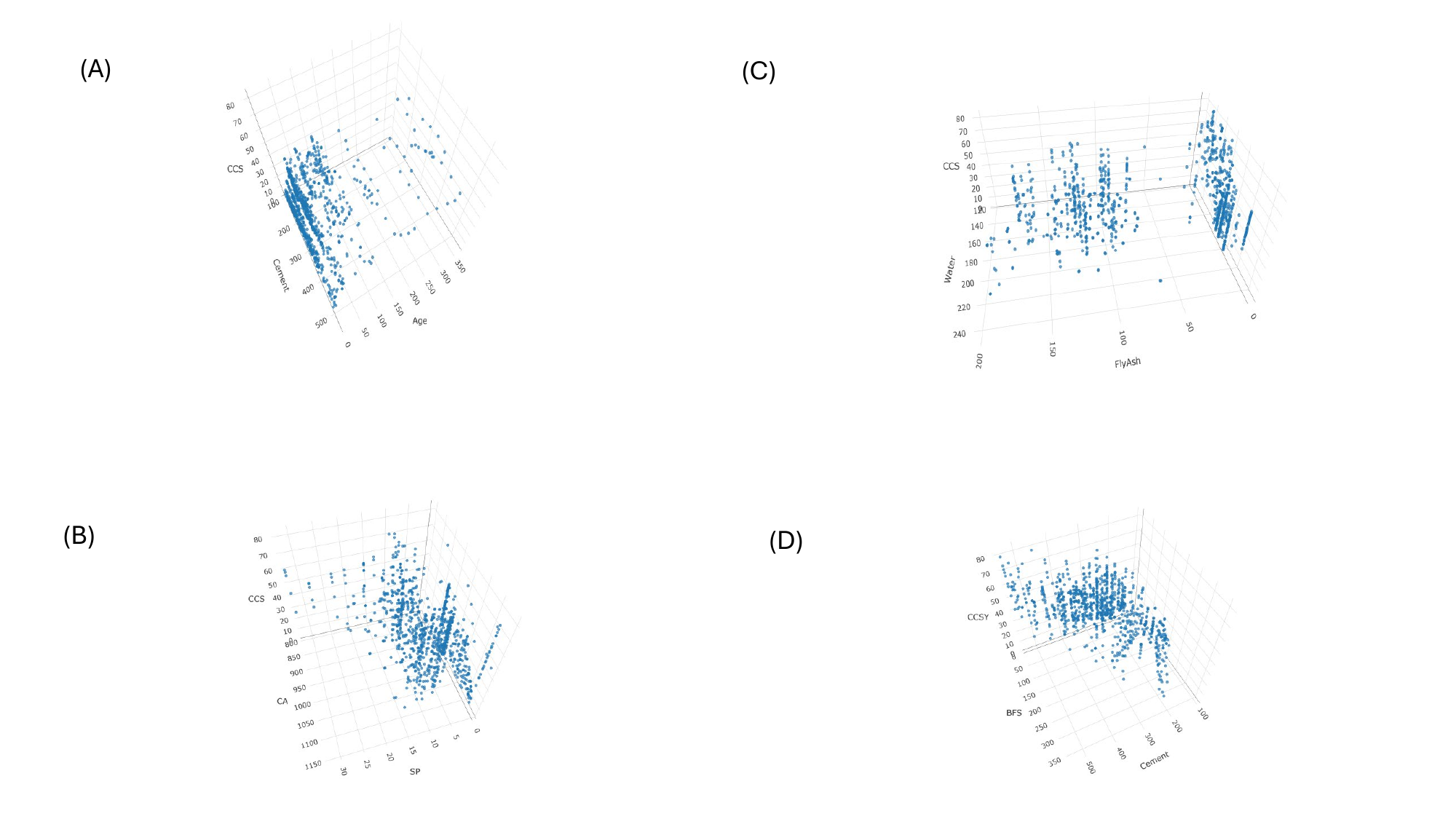}
 \caption{Complex nonlinear relationships between CCS and the 8 covariate features. }
 \label{CCSrelations}
 \end{figure}

The CCS data set consists of 1030 data points of 9 dimensions. The CCS measurement is taken as the 1D response variable, while the remaining 8 variables are taken as eight 1D quantitative covariate features: Cement, Age, Fly Ash, Blast Furnace Slag, Water, Superplasticizer, Fine Aggregate, and Coarse Aggregate. The complex relationships between CCS and the 8 covariate features can be seen through the four snapshots of 3D plots shown in the four panels of Fig.~\ref{CCSrelations}.

In panel (A) and (D), the order-2 relationships between CCS and (Cement, Age) and between CCS and (Cement, Blast Furnace Slag) are very difficult to describe or even visualize. Across various values of Age and Blast Furnace Slag, respectively, a series of scatter plots with varying and unfixed patterns are seen. Similar impressions result across the cross-sectional plots with respect to different values of Cement. In the panels (B) and (C), the order-2 relationships between CCS and (Coarse Aggregate, Superplasticizer) and between CCS and (Water,Fly Ash) are rather messy without any clues of patterns. One catching phrase seems to capture the order-2 interacting effects of covariate onto CCS: Complex nonlinearity.

\begin{figure}[ht!]
 \centering
 \includegraphics[width=1.0\textwidth]{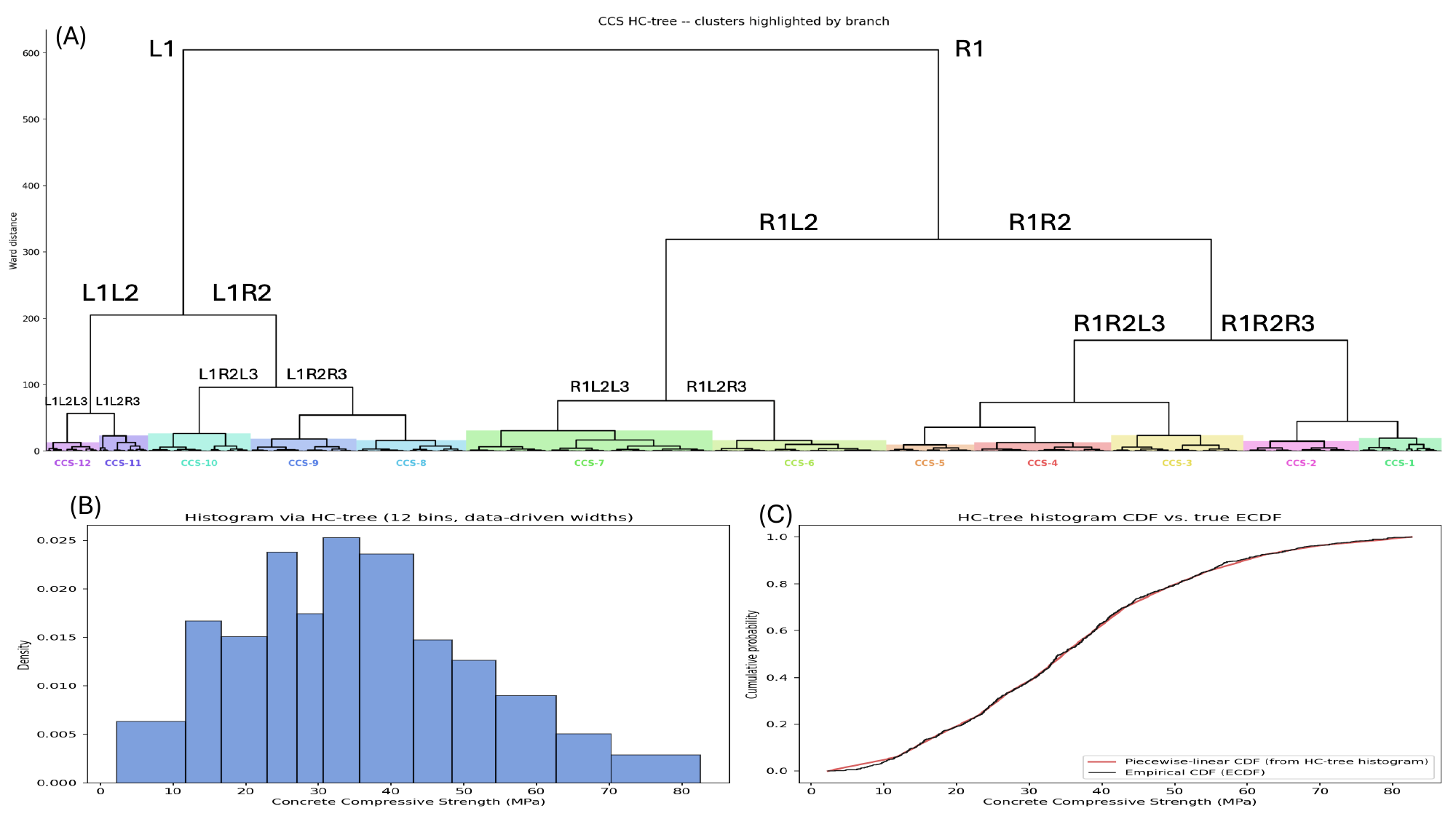}
 \caption{Distributional information of CCS: (A) Hierarchical Clustering tree of CCS with 12 color-encoded cluster-nodes; (B) histogram with 12 bins; (C) the histogram's distribution as a piecewise linear approximation of ECDF. }
 \label{CCSHCtree}
 \end{figure}

With these complex nonlinear relationships in mind, the task of exploring the CCS system's dynamics seems rather challenging. In this paper, we employ CT to tackle this challenge. The first task of CT is to build a Taxonomic Hierarchy. Since CCS is a 1D quantitative variable, it is natural to take a Hierarchical Clustering (HC) tree of CCS as the starting basis. An HC-tree of CCS is constructed using the Euclidean distance with the Ward-D2 module, as shown in panel (A) of Fig.~\ref{CCSHCtree}. A cut at the 12-cluster level of this HC-tree gives rise to a histogram with 12 bins \cite{FR2018}, as shown in panel (B). In panel (C), we demonstrate that the distribution of this histogram is a piecewise-linear approximation of the Empirical Cumulative Distribution Function (ECDF). That is, this histogram has captured the distributional information of the 1030 CCS measurements at both global and local scales rather well. Further, given the known randomness within all bins, this histogram is turned into a seemingly informative, categorized CCS response variable.

Nonetheless, in panel (C) of Fig.~\ref{CCSHCtree}, we see a degree of discrepancy between the first linear piece and the ECDF curve over the range of the first bin. This discrepancy indicates the need to determine whether a further split of the first bin into two sub-bins can be justified. Here, computational justification means checking whether the CCS measurements belonging to these two sub-bins can be separated using the information provided by their 8 covariate feature measurements. In terms of the CCS system's dynamics, if this computational justification for separating the two sub-bins holds, then they are understood to have distinct local CCS dynamics; if the justification fails, the two sub-bins are homogeneous within the local system dynamics of the first bin, with respect to the available finite sample data.

By carrying the same argument across all branches, including the 12 clusters, we systematically confirm or dispute each branch-split along the tree-branching process, starting from the top internal node of the HC-tree in panel (A) of Fig.~\ref{CCSHCtree}. This is the task of constructing the Taxonomic Hierarchy.

We apply the Left(L)-and-Right(R) encoding scheme throughout the branching process of the CCS HC-tree, as shown in panel (A) of Fig.~\ref{CCSHCtree}. The first step of Taxonomic Hierarchy construction is to confirm or dispute the separation of L1-vs-R1. For this task, we set up a binary response variable with two response (Re) categories, $\{L1, R1\}$, and categorize all covariate (Co) feature-sets of all orders with respect to their HC-tree individually. That is, all members of a feature-set are fused using a Hierarchical Clustering algorithm via an HC-tree, and a cut on this HC-tree provides the categorization of the feature-set. This is how we accommodate interacting effects of any order. Here, the order refers to the number of members in a feature-set.

Suppose this categorized covariate feature-set has 5 categories. Then the associative relationship between this binary response variable and the covariate feature-set is summarized by a $5\times 2$ contingency table. Each of the 5 rows provides a piece of potential associative information through the conditional distribution of the binary response variable, equipped with its own finite-sample precision provided by the visible randomness of the contingency table. This is the foundation for performing Scientific Data Analysis (SDA) to explore all potential pieces of associative information within the Re-Co dynamics; see details of SDA developments and illustrations in \cite{CTonIris,CTonpenguin,omotayo}. Each piece of associative information passing a reliability check is termed a major feature-category, with an order equal to the number of features involved.

A 3-digit coding scheme, defined as $a\times 2^0+ b\times 2^1 +c\times 2^2$ for a binary triplet representing the presence or absence of three covariate features, is employed for each major feature-category. With $I_{ABC}$ being an indicator function, the first digit is calculated corresponding to $(a, b, c)=(I_{Cement}, I_{Age}, I_{Fly Ash})$, the second digit to $(a, b, c)=(I_{Blast Furnace Slag}, I_{Water}, I_{Superplasticizer})$, and the third digit to $(a,b)=(I_{Fine Aggregate}, I_{Coarse Aggregate})$. For instance, the code 123 stands for the feature-set $\{Cement, Water, Fine Aggregate, Coarse Aggregate\}$.

\subsection{1st layer of the Taxonomic Hierarchy}
\begin{figure}[ht!]
 \centering
 \includegraphics[width=1.0\textwidth]{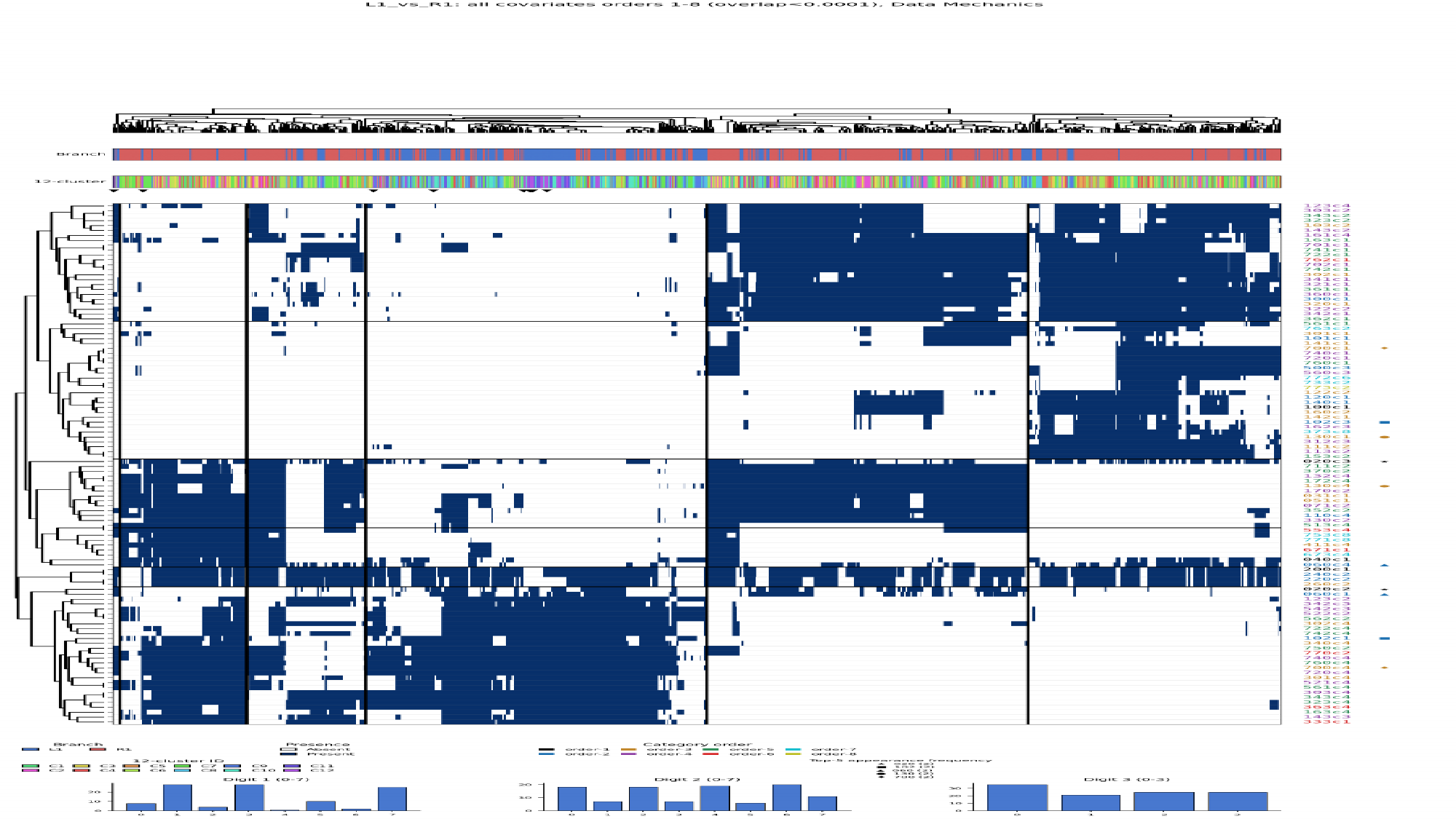}
 \caption{Heatmap for L1-vs-R1. }
 \label{L1vsR1}
 \end{figure}

Via SDA on the Re-Co dynamic at the top internal node of HC-tree in panel (A) of Fig.~\ref{CCSHCtree}, all computed and confirmed pieces of associative information are collected and displayed in a heatmap, as shown in Fig.~\ref{L1vsR1}. Here, a heatmap is a binary bipartite network with all study-subjects arranged along the column-axis, while each selected major feature-category -- represented as a binary row-vector of presence or absence across study-subjects -- is arranged along the row-axis. After iteratively applying the HC algorithm with a modified Euclidean distance, with respect to clustering structures found on the row- and column-axis in the previous iteration -- an operational computing algorithm on the matrix lattice called Data Mechanics (DM) \cite{CF12,FC14} -- the heatmap manifests interacting relational block-patterns, framed by clusters of study-subjects on the column-axis and clusters of major feature-categories of various orders on the row-axis.

The 3-digit code column on the right-hand side of the heatmap displays the full names of all major feature-categories of various orders. The frequency histograms of the three digits are displayed below the heatmap. The histogram of the 1st digit shows three evident bars at $\{1, 3, 7\}$, meaning that $Cement$, $\{Cement, Age\}$, and $\{Cement, Age, Fly Ash\}$ appear almost equally often across the entire column-list of major feature-categories. In contrast, $\{0, 2, 4, 6\}$ appear almost equally often in the 2nd digit across the column-list of names, meaning that none of $(Blast Furnace Slag, Water, Superplasticizer)$, $Water$ alone, $Superplasticizer$ alone, or $\{Water, Superplasticizer\}$ appear often within the major feature-categories. For the 3rd digit, $0$ has the highest frequency, meaning that neither $Fine Aggregate$ nor $Coarse Aggregate$ appears often.

After applying DM for three rounds, the resultant heatmap is sustained by a $6\times 6$ array of blocks, some nearly full of 1's, some nearly full of 0's. Each block is framed by one cluster on the row-axis and one cluster on the column-axis. The six clusters on the column-axis are either primarily red-color-encoded, for study-subjects coming from the R1 branch of the CCS HC-tree, or primarily blue-color-encoded, for study-subjects belonging to the L1 branch. These clear separations, though not $100\%$, strongly indicate that the two major branches, L1 and R1, are highly separable. This decision confirms the 1st layer of the Taxonomic Hierarchy.

It is worth noting that each of these six clusters of study-subjects is also characterized by a vertical block-chain within the heatmap, since each block of 1's signals a strong, locality-based mechanistic dependence among all involved major feature-categories, arising from sharing the same cluster of study-subjects. Therefore, each vertical block-chain serves as a basis for interpreting and explaining the cluster of study-subjects through the presence or absence of mechanistic dependence of various kinds.

\subsection{2nd layer of the Taxonomic Hierarchy}
Having confirmed the separation of L1 against R1 at the 1st layer of the Taxonomic Hierarchy, we proceed to explore the possible structural heterogeneity on the 2nd layer. This layer is defined by two questions: Should L1 be split into two sub-branches, L1L2 and L1R2? And should R1 likewise be split into R1L2 and R1R2?

\begin{figure}[ht!]
 \centering
 \includegraphics[width=1.0\textwidth]{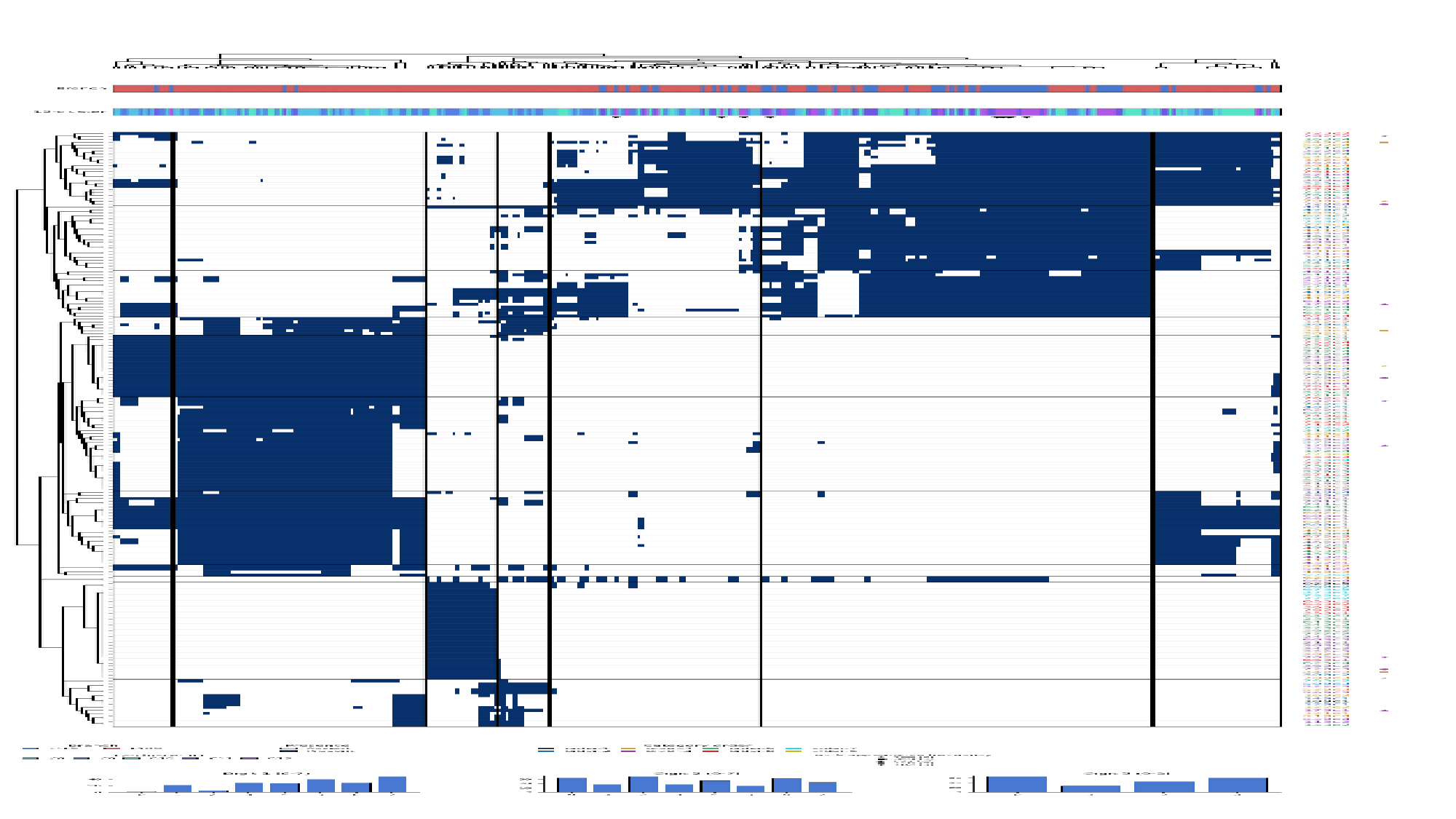}
 \caption{Heatmap for L1L2-vs-L1R2. }
 \label{L1L2vsL1R2}
 \end{figure}

If L1 is split into L1L2 and L1R2, a binary response variable is defined, and so is the Re-Co dynamic on the subset of data belonging to branch L1. Correspondingly, SDA is applied and its resultant heatmap is constructed, as shown in Fig.~\ref{L1L2vsL1R2}. The separation between L1L2 and L1R2 is evidently confirmed.

\begin{figure}[ht!]
 \centering
 \includegraphics[width=0.8\textwidth]{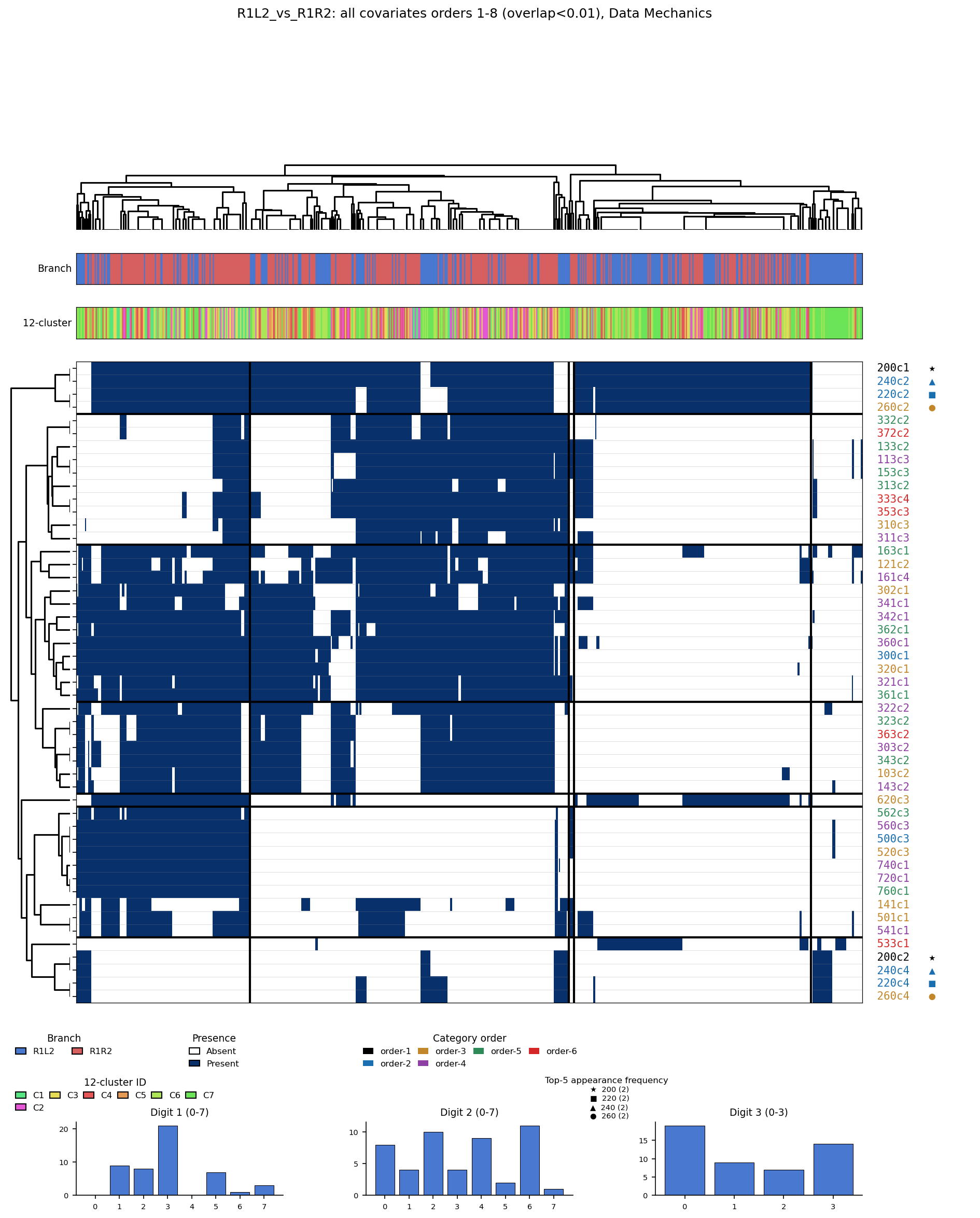}
 \caption{Heatmap for R1L2-vs-R1R2. }
 \label{R1L2vsL1R2}
 \end{figure}

\begin{figure}[ht!]
 \centering
\includegraphics[width=1.0\textwidth]{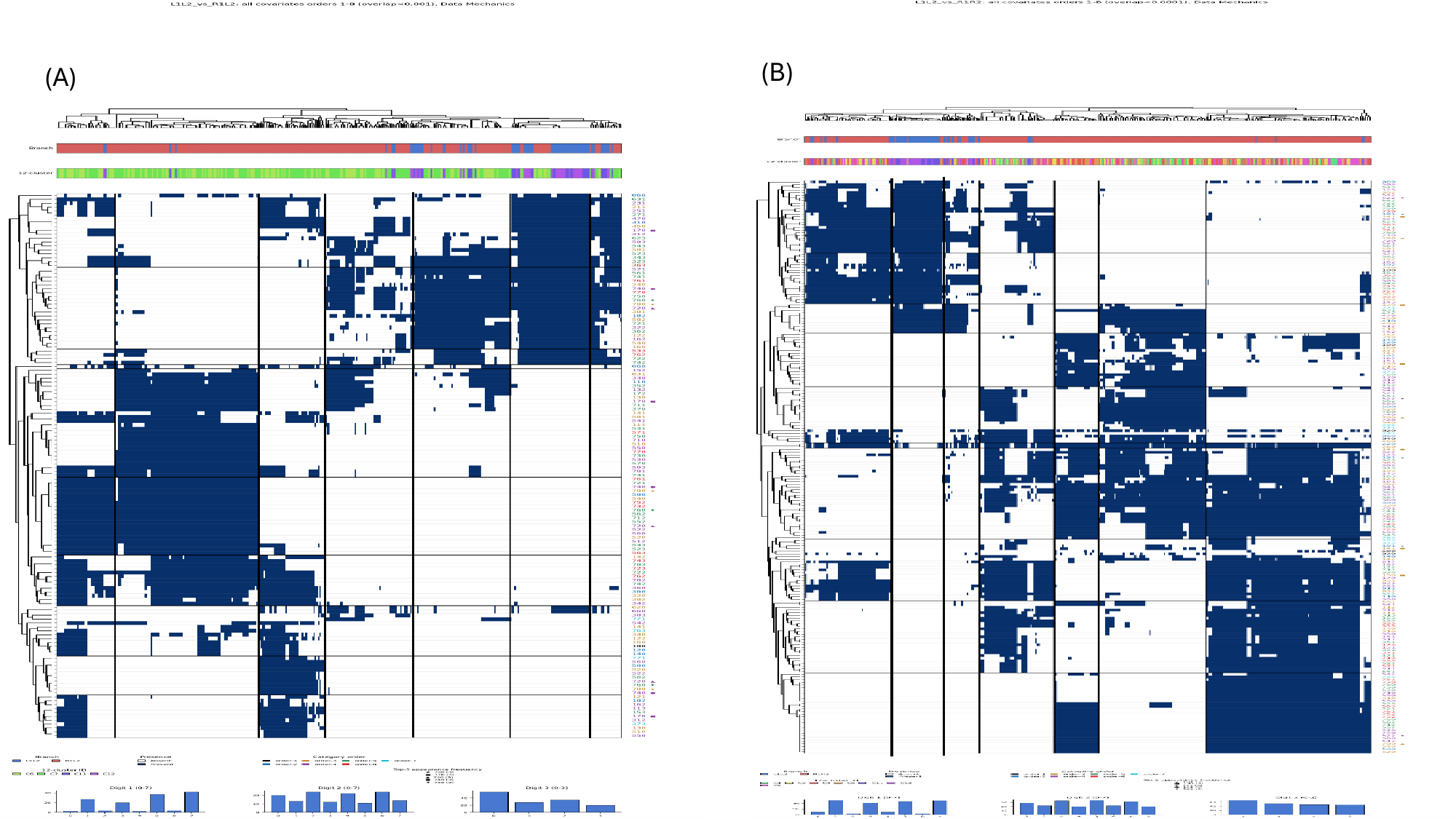}
 \caption{Heatmaps for L1L2-vs-R1L2 and L1L2-vs-R1R2. }
 \label{R1L2nL1R2vsL1L2}
 \end{figure}

The SDA-based heatmap for separating R1L2 and R1R2 is less evident, but still visible, as shown in Fig.~\ref{R1L2vsL1R2}. We provide additional evidence to support this separation. By comparing two heatmaps pertaining to R1L2-vs-L1L2 and R1R2-vs-L1L2, shown in Fig.~\ref{R1L2nL1R2vsL1L2}, this pair of SDA-based heatmap comparisons shed further, distinct light on the difference between R1L2 and R1R2. Similar supporting information can be derived by comparing two heatmaps pertaining to R1L2-vs-L1R2 and R1R2-vs-L1R2.
% shown in Fig.~\ref{R1L2nL1R2vsL1R2}.

%   \begin{figure}[h!]
% \centering
%\includegraphics[width=1.0\textwidth]{CCS-L1R2-vs-R1L2-n-R1R2-new.pdf}
% \caption{Heatmaps for L1R2-vs-R1L2 and L1R2-vs-R1R2. }
% \label{R1L2nL1R2vsL1R2}
% \end{figure}

\subsection{3rd layer of the Taxonomic Hierarchy}
Having confirmed branches L1 and R1 at the 1st layer, both split into two sub-branches to form the four branches L1L2, L1R2, R1L2, and R1R2 at the 2nd layer. Proceeding to the 3rd layer, each of the four branches is split into two sub-branches, and the resultant splits are confirmed using the same principle described in the Introduction and used at the 1st and 2nd layers.

Since a split on a data subset creates a Re-Co dynamic with a binary response variable coupled with a collection of covariate features, any computed and confirmed piece of associative information is evidence, passing a reliability check, for heterogeneity. When multiple pieces of associative information share the same cluster of study-subjects, the evidence for heterogeneity is substantially strengthened -- that is, strong evidence of heterogeneity appears as a heatmap sustained by multiple clear, solid block-chains across multiple clusters of study-subjects.

\begin{figure}[ht!]
 \centering
\includegraphics[width=1.0\textwidth]{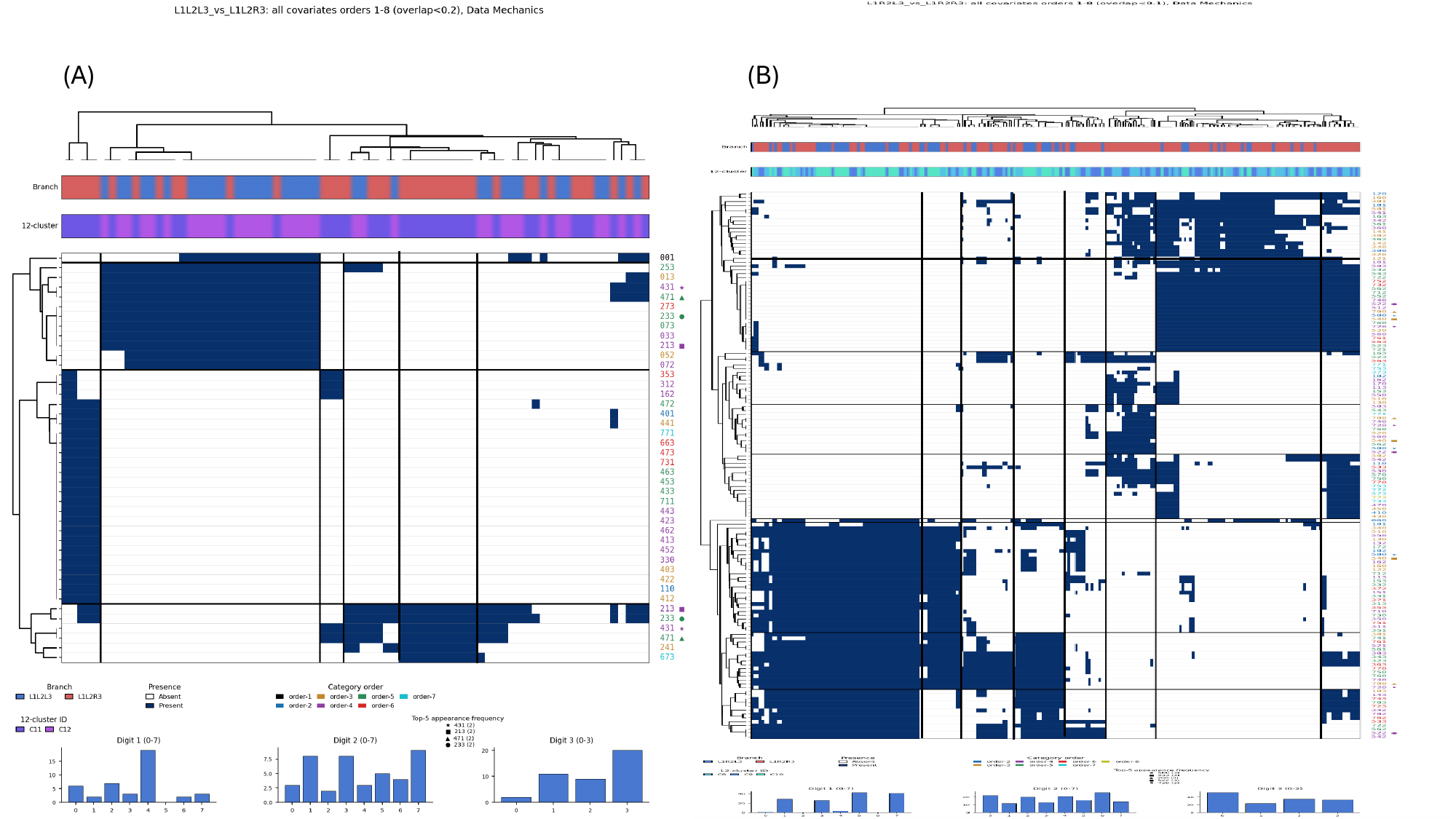}
 \caption{Heatmaps for (A) L1L2L3-vs-L1L2R3 and (B) L1R2L3-vs-L1R2R3. }
 \label{splitL1L2nL1R2}
 \end{figure}

\begin{figure}[ht!]
 \centering
\includegraphics[width=1.0\textwidth]{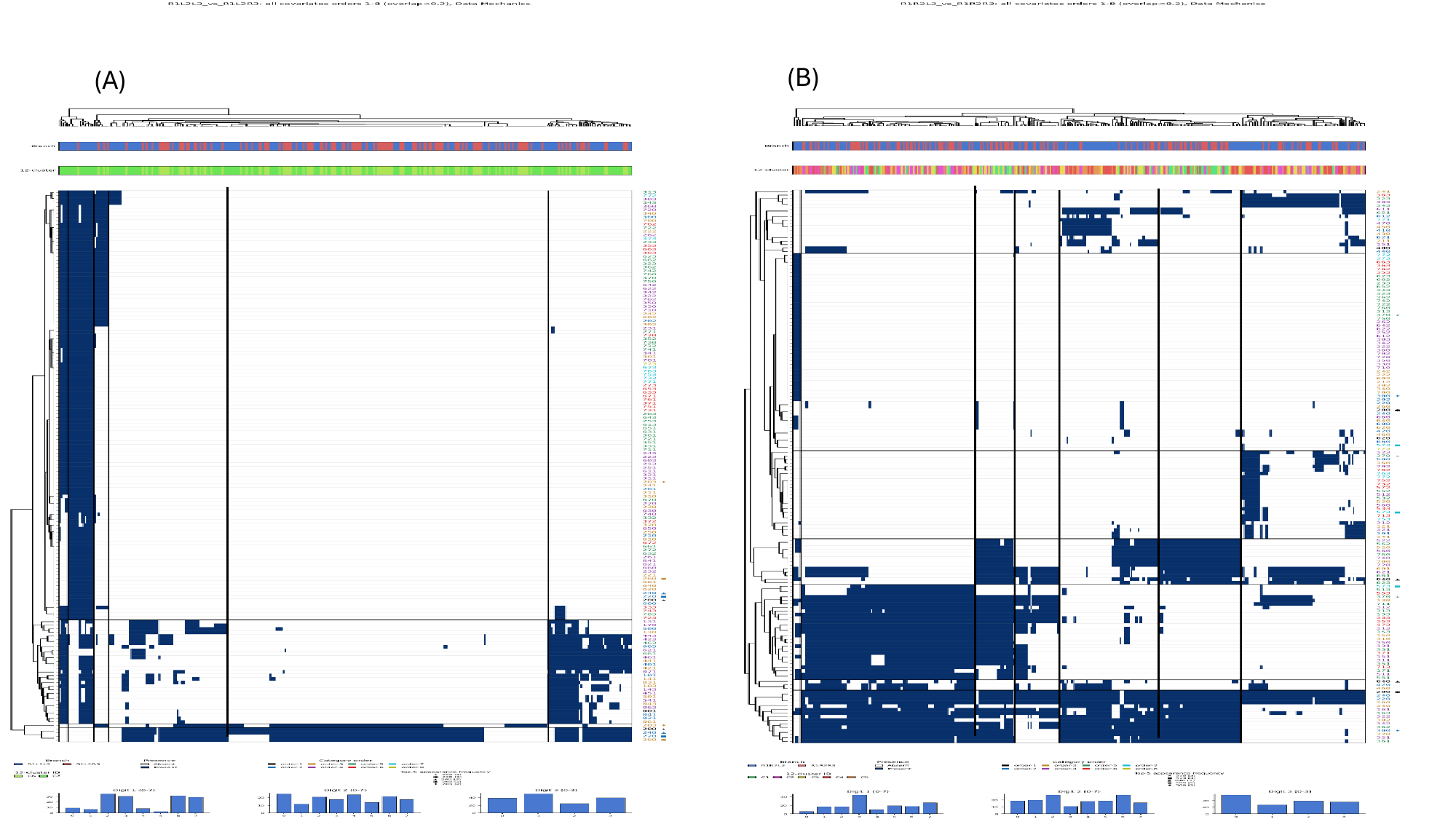}
 \caption{Heatmaps for (A) R1L2L3-vs-R1L2R3 and (B) R1R2L3-vs-R1R2R3. }
 \label{splitL1R2nR1R2}
 \end{figure}

The two panels of Fig.~\ref{splitL1L2nL1R2} show the two resultant heatmaps from splitting branches L1L2 and L1R2, respectively. In panel (A), there are three nearly solid blocks of 1's, each with a primary membership of L1L2L3 and L1L2R3. We therefore confirm the split of L1L2 into two sub-branches, L1L2L3 and L1L2R3. Likewise, we confirm the split of branch L1R2 into L1R2L3 and L1R2R3.
Likewise, we confirm the split of R1L2 into R1L2L3 and R1L2R3 via the heatmap shown in panel (A) of Fig.~\ref{splitL1R2nR1R2}, and the split of R1R2 into R1R2L3 and R1R2R3 via the heatmap in panel (B).

\subsection{4th layer of the Taxonomic Hierarchy}
\begin{figure}[ht!]
 \centering
\includegraphics[width=1.0\textwidth]{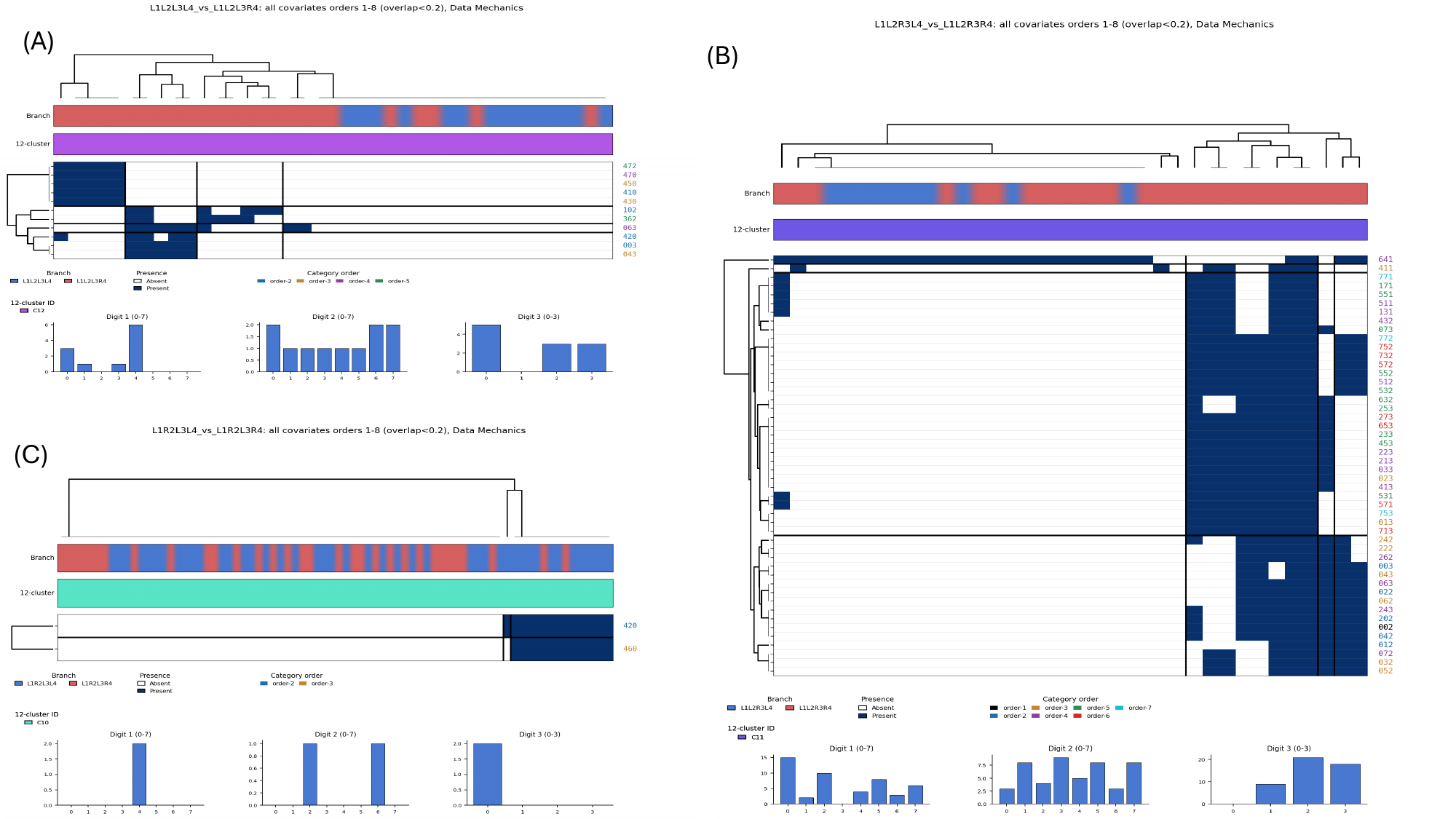}
 \caption{Three heatmaps exploring the potential for splitting: (A) L1L2L3L4-vs-L1L2L3R4; (B) L1L2R3L4-vs-L1L2R3R4; (C) L1R2R3L4-vs-L1R2R3R4. }
 \label{4thlayeratL1}
 \end{figure}

At the 4th layer, based on SDA computations and the heatmap representations shown in the three panels of Fig.~\ref{4thlayeratL1}, we decide to split the two sub-branches of L1L2, while keeping the two sub-branches of L1R2 intact without further splitting. That is, L1L2L3 is split into L1L2L3L4 (marked as cluster-12L) and L1L2L3R4 (marked as cluster-12R); CCS measurements in cluster-12R are larger than those in cluster-12L. The branch L1L2R3 is split into L1L2R3L4 (marked as cluster-11L) and L1L2R3R4 (marked as cluster-11R); CCS measurements in cluster-11R are less than those in cluster-11L. The branches L1R2L3 (marked as cluster-10) and L1R2R3 (marked as cluster-$8\&9$) are kept intact, as no confirmed associative information pieces were found in the SDA computations for splitting L1R2R3. These cluster numbers refer to Fig.~\ref{CCSHCtree}.

\begin{figure}[ht!]
 \centering
\includegraphics[width=1.0\textwidth]{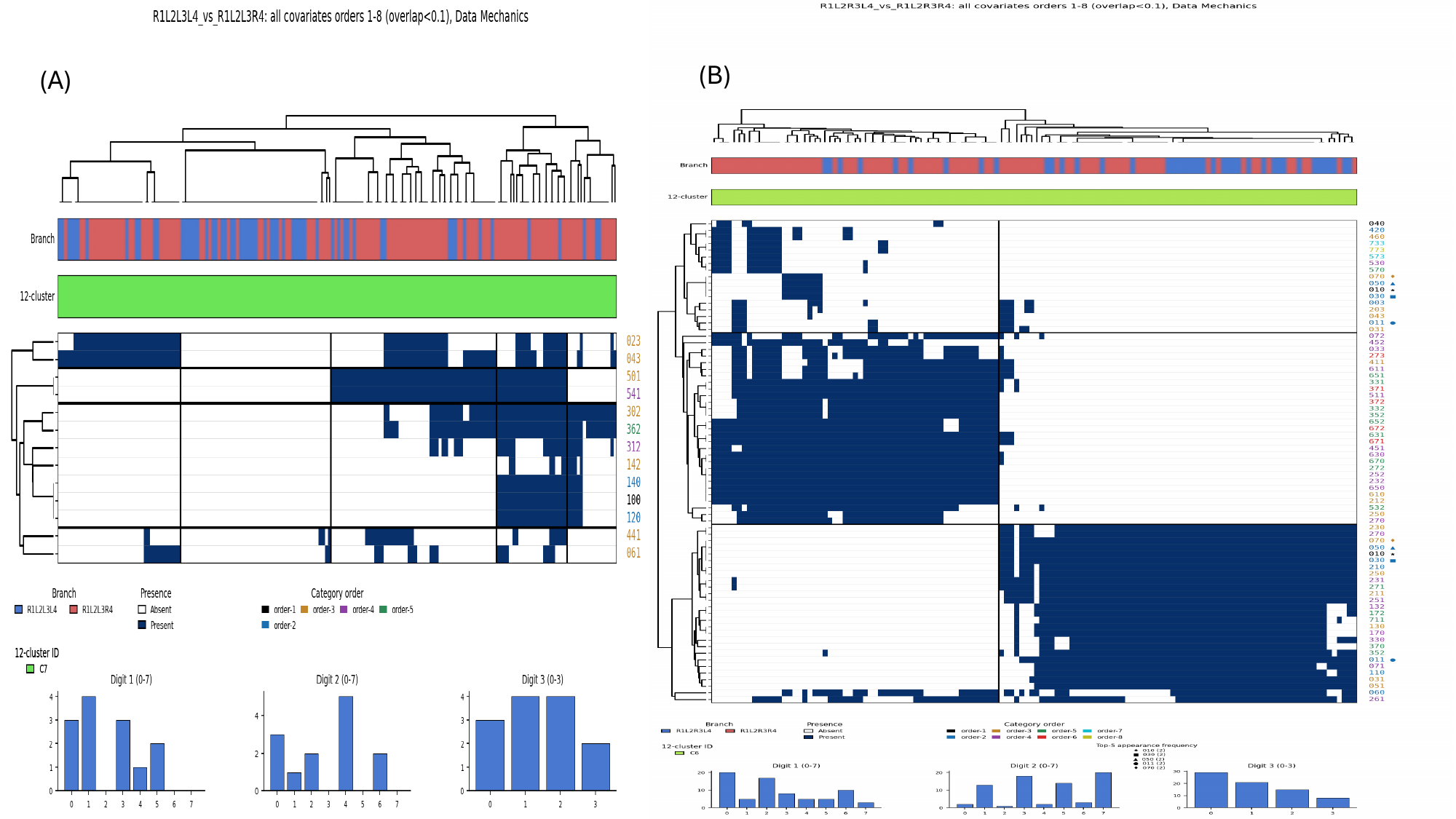}
 \caption{Two heatmaps exploring the potential for splitting: (A) R1L2L3L4-vs-R1L2L3R4; (B) R1L2R3L4-vs-R1L2R3R4. }
 \label{4thlayeratR1L2}
 \end{figure}

Next, we consider the four branches constituting branch R1. As shown in the two panels of Fig.~\ref{4thlayeratR1L2}, branch R1L2L3 (marked as cluster-7) is split into R1L2L3L4 (marked as cluster-7L) and R1L2L3R4 (marked as cluster-7R). Likewise, branch R1L2R3 (marked as cluster-6) is split into R1L2R3L4 (marked as cluster-6L) and R1L2R3R4 (marked as cluster-6R).

\begin{figure}[ht!]
 \centering
\includegraphics[width=1.0\textwidth]{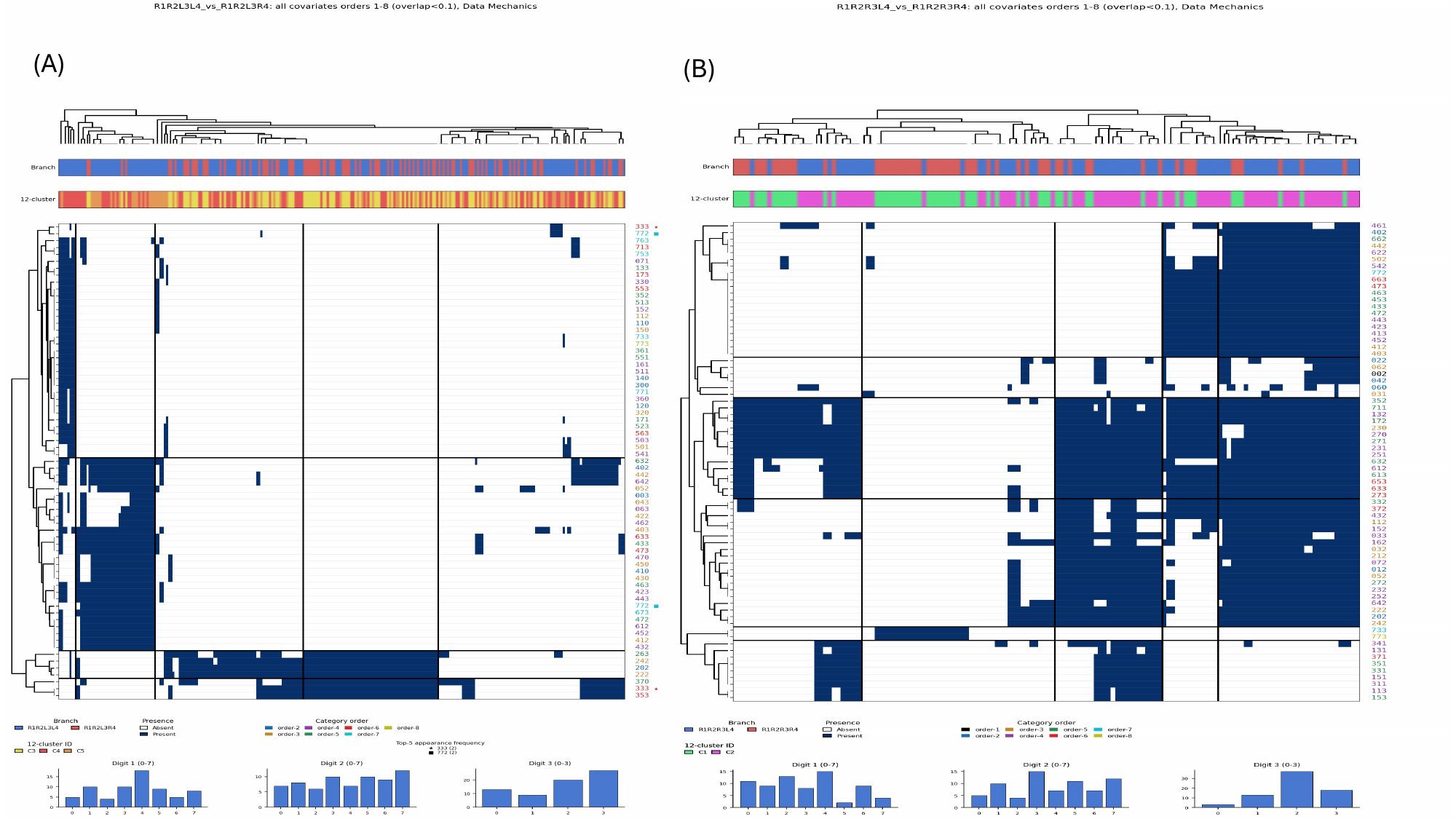}
 \caption{Two heatmaps exploring the potential for splitting: (A) R1R2L3L4-vs-R1R2L3R4; (B) R1R2R3L4-vs-R1R2R3R4. }
 \label{4thlayeratR1R2}
 \end{figure}

As shown in the two panels of Fig.~\ref{4thlayeratR1R2}, branch R1R2L3 is split into R1R2L3L4 (marked as cluster-$4\&5$) and R1R2L3R4 (marked as cluster-3). Branch R1R2R3 is split into R1R2R3L4 and R1R2R3R4 (marked as cluster-2 and cluster-1, respectively).

\begin{figure}[ht!]
 \centering
\includegraphics[width=1.0\textwidth]{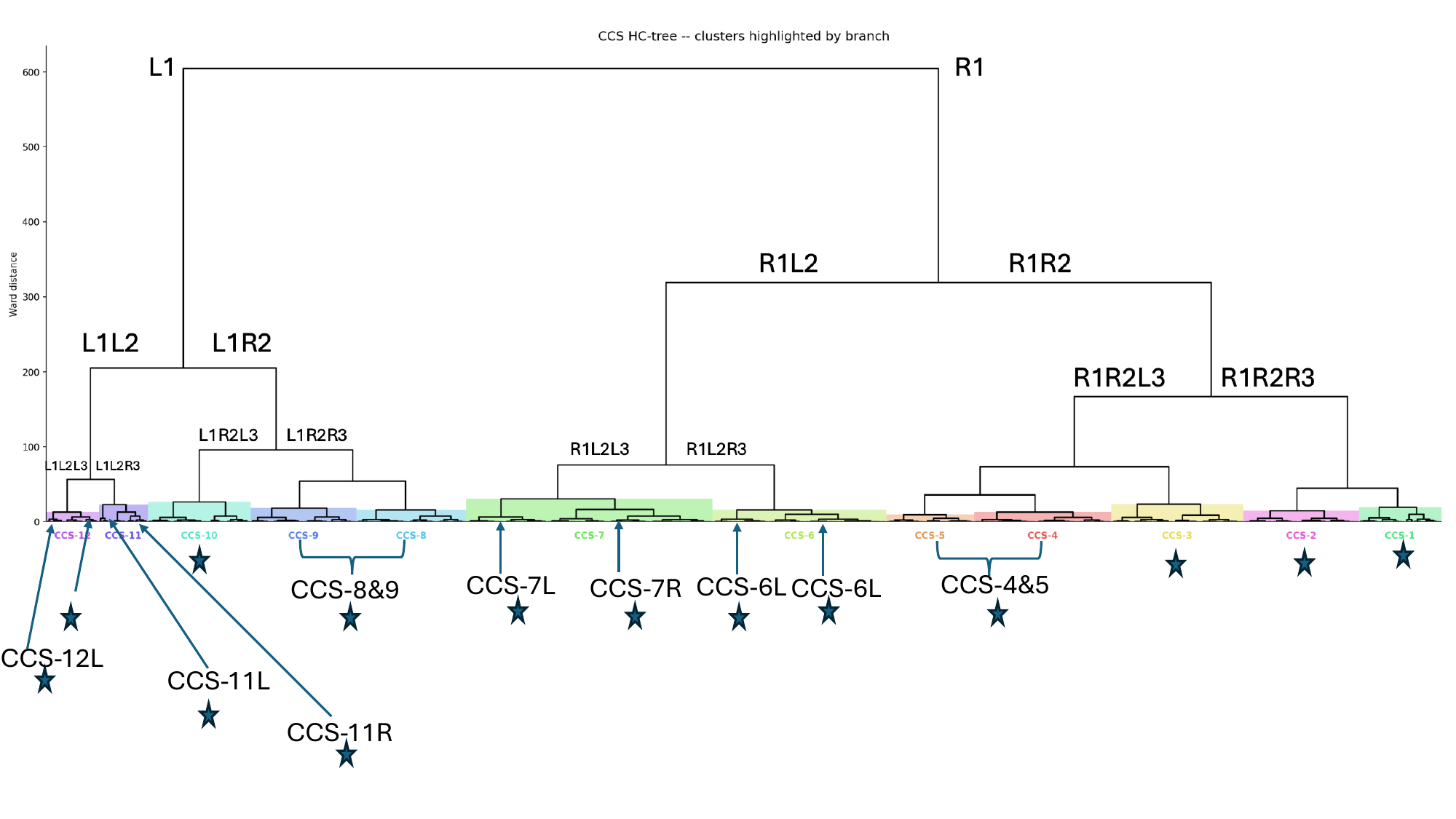}
 \caption{Taxonomic Hierarchy with all ending nodes marked with $\star$.}
 \label{CCSTH}
 \end{figure}

The resultant Taxonomic Hierarchy is shown in Fig.~\ref{CCSTH}, with all ending nodes marked with $\star$. Each ending node is taken as a class where the response homogeneity is found being coherent with covariate homogeneity.

\section{Randomness or Nonlinearity?}
The CCS system's dynamics are known for their hardly describable nonlinearity and randomness in the field of civil engineering \cite{yeh}. What does nonlinearity mean here? How does randomness within the CCS system vary with quantitative changes in 1D CCS? No explicit answers to these two questions can be found in the literature. Nevertheless, since we have seen high-order interacting effects when constructing the Taxonomic Hierarchy in the previous section, realistic, if implicit, answers can be deduced to a great extent, given that such interacting effects only have unknown functional forms. Consequently, any assumptions about functional formulas for CCS would be artificial or even unscientific; likewise, any assumed distributional randomness would be unrealistic, or even elusive.

In this section, we explicitly reveal nonlinearity and randomness directly based on the Taxonomic Hierarchy. These two chief intrinsic characteristics of the CCS system become visible by reiterating one fundamental character, or requirement, of the Taxonomic Hierarchy. Seeing these two characteristics explicitly helps scientists peek into the CCS complex system's dynamics, especially through their finite-sample nature. Indeed, nonlinearity and randomness are two chief intrinsic characteristics of most real-world complex systems. If we can see these two characteristics within the CCS system, we can likewise see them in any other complex system, for a better understanding.

The Taxonomic Hierarchy is, in essence, a basis for exploring the multiscale nonlinearity and randomness embraced within the CCS system's dynamics, because of its characterization as a heterogeneity-vs-homogeneity map, as discussed in the previous section. The concept of randomness is already embraced within a class, since a class is an established locality of the CCS system where coherence of similarity-based homogeneity between the response feature and covariate features is confirmed. This operational concept of randomness both promoted and required our protocol for building the Taxonomic Hierarchy, via progressive splitting decisions on each parent branch by comparing its two sub-branches. Such comparisons are carried out based on SDA-computed associative information pieces regarding the corresponding Re-Co dynamic. That is, randomness of finite-sample nature within the CCS complex system is seen and understood through locality-specific coherence between the 1D response CCS and the 8D covariate-features' ``homogeneity.'' As such, this randomness-based collection of classes partitions the entire CCS system, allowing us to see the CCS system's dynamics from the locality-randomness perspective. This is one defining merit of Computational Taxonomy, with significant implications for DoE.

In contrast, as another defining characteristic of Computational Taxonomy, the concept of heterogeneity among all recognized classes collectively points to the fact that the response's structural discrepancy must be interpretable via covariate features. In particular, when two non-neighboring classes' observable differences in CCS measurements cannot be perfectly separated and interpreted by covariate features, one aspect of the CCS system's scale-sensitive nonlinearity is found and confirmed as being embedded within heterogeneity. That is, nonlinearity is seen through incoherence between the response feature's and the covariate features' ``locality-specific homogeneity.'' Such incoherence is especially evident through the mixing of CCS measurements from two distinct response localities. As such, the CCS system's nonlinearity, at various scales, can and must be explored by comparing two branches located on two distinct branching paths of the Taxonomic Hierarchy. This is a brand-new, rigorous definition of nonlinearity that does not refer to any unknown hypothetical functional form -- another crucial merit of the Taxonomic Hierarchy, with significant implications for DoE.

Exploring randomness and nonlinearity is not only necessary, but crucial for understanding the CCS system's dynamics. The importance of figuring out randomness embedded within data was emphasized by John Tukey \cite{tukey} and A. N. Kolmogorov \cite{kolmogorov} many decades ago. It is intuitive and expected that randomness and nonlinearity are highly intertwined, globally and locally, with distinct degrees, in this CCS data set. In this section, we demonstrate their interlocked waxing-and-waning patterns of mixing on CCS measurements across the tree-branches of the Taxonomic Hierarchy.

A glimpse of such interlocking patterns can be seen as follows. As reported in the three panels of Fig.~\ref{CCSHCtree}, the three aspects of variation-based patterns regarding CCS measurements alone do not automatically account for the CCS system's randomness. Such variations in CCS measurements must be checked with respect to all possible covariate feature-sets, to confirm which part is randomness and which part is not. On the other hand, the system's nonlinearity should ideally be checked across the entire expansion of the Taxonomic Hierarchy, in order to determine its scope within the CCS system's dynamics. Both checking tasks, corresponding to varying Re-Co dynamics at varying scales, are performed by applying SDA computations. All SDA-based pieces of associative information are displayed on the heatmap platform. Some heatmaps have clear block-structures; some do not. On those heatmaps with clear block-structures, we look for any block framed by a cluster of study-subjects with mixed IDs. Such mixing of study-subject IDs from two non-neighboring classes clearly indicates incoherence between the locality-specific homogeneity of covariate features and the locality-specific homogeneity of the response -- evidence of nonlinearity. If such mixing patterns are seen between two neighboring classes sharing a common parent branch, then randomness and nonlinearity mingle and become difficult to distinguish. This fact contributes to the difficulties facing DoE in particular, as seen in the next section.

In summary, the Taxonomic Hierarchy provides a basis for exploring the mixing patterns and intertwined relations between these two intrinsic characteristics -- randomness and nonlinearity -- which are theoretically quite distinct, but realistically very difficult to differentiate well in some settings. Ideally, when comparing two far-apart branches, nonlinearity has more potential than randomness to be seen through mixing patterns. When comparing two nearby branches, randomness is more likely to be the chief factor driving the visible mixing patterns. Between these two extremes, nonlinearity and randomness mix, with evolving degrees, in a waxing-and-waning fashion; they are difficult to distinguish perfectly. Through the above descriptions of these two concepts, the Taxonomic Hierarchy is illustrated below to give rise to an evolution of mixing patterns, from far-apart to nearby branch-vs-branch comparisons.

\begin{figure}[h!]
 \centering
\includegraphics[width=1.0\textwidth]{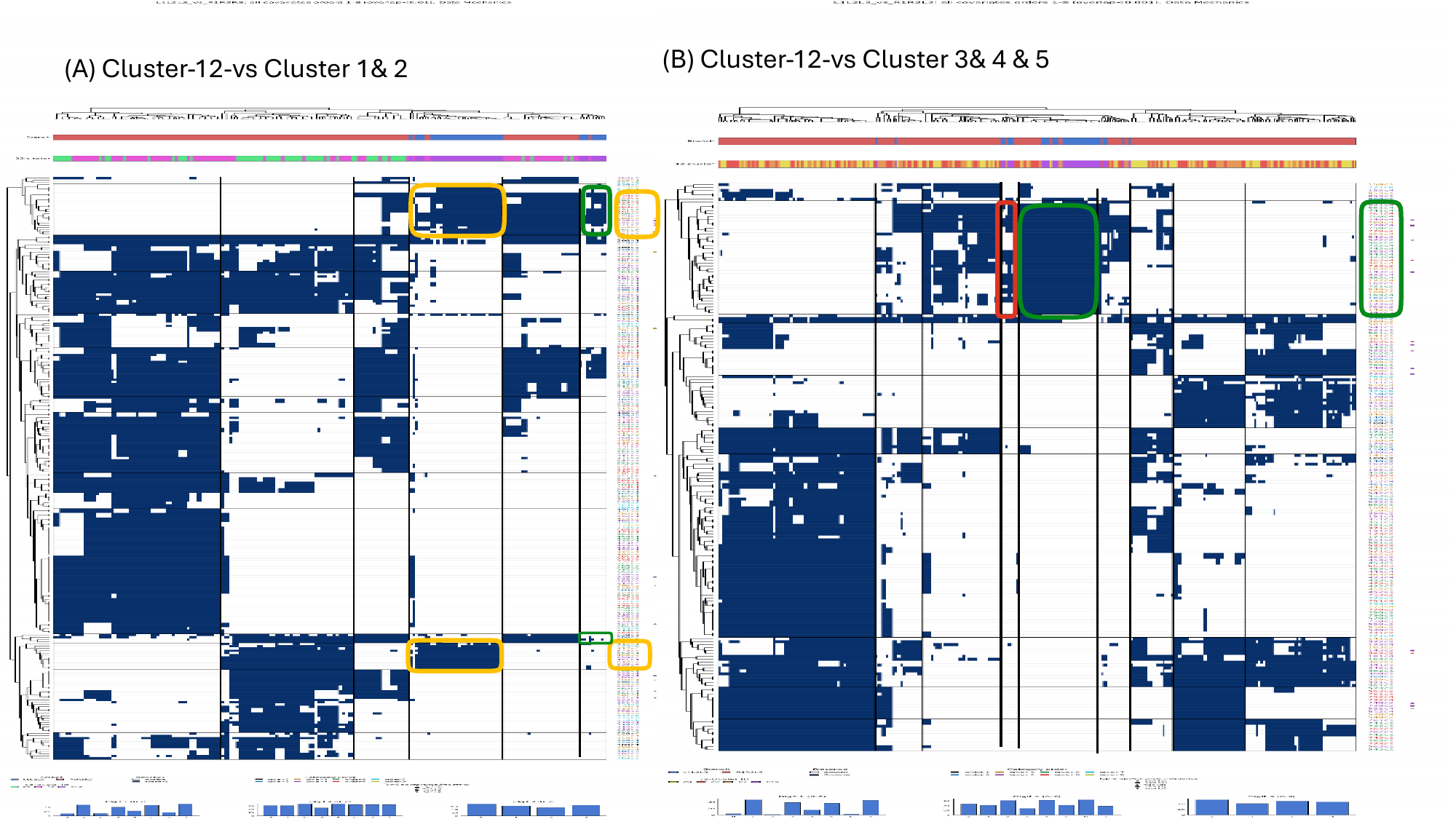}
 \caption{Two heatmaps exploring the potential for randomness or nonlinearity: (A) L1L2L3-vs-R1R2R3; (B) L1L2L3-vs-R1R2L3. }
 \label{cluster12vs1n2}
 \end{figure}

Consider the branch-vs-branch comparison of Cluster-12, the farthest-right bin, encoded as L1L2L3, collecting the largest values among the 12 clusters of CCS measurements, against Cluster-$1\&2$, the union of the first and second bins on the farthest left of the histogram, encoded as R1R2R3, with CCS values significantly smaller than those in Cluster-12. The heatmap for this comparison is shown in panel (A) of Fig.~\ref{cluster12vs1n2}. Upon its HC-tree, denoted $HC[12$-$vs$-$1\&2]$, on the column-axis, its left branch, denoted $HC[12$-$vs$-$1\&2](L1)$, consists exclusively of members from Cluster-$1\&2$. This is not surprising; it is even expected. Upon its right branch, denoted $HC[12$-$vs$-$1\&2](R1)$, we also see that its sub-branches -- $HC[12$-$vs$-$1\&2](R1L2)$, $HC[12$-$vs$-$1\&2](R1R2L3)$, $HC[12$-$vs$-$1\&2](R1R2R3L4)$, and $HC[12$-$vs$-$1\&2](R1R2R3R4L5)$ -- contain members from Cluster-$1\&2$ exclusively as well. However, two circled sub-branches of branch $HC[12$-$vs$-$1\&2](R1R2R3R4R5)$ contain mixed members from both Cluster-$1\&2$ and Cluster-12.

This observation is primarily attributed to nonlinearity. The mixing appears because members from two rather far-apart CCS-based branches share very similar binary column-vectors, which are covariate-feature-set-based characteristics. Such a binary column-vector is called a study-subject's individual character-landscape within a heatmap. Two study-subjects can have nearly equal individual character-landscapes yet rather distinct CCS values. This phenomenon is unlikely to be caused by randomness. Nonetheless, this striking nonlinearity-driven mixing clearly indicates that the CCS system is highly non-smooth and nearly impossible to predict precisely.

Likewise, in panel (B) of Fig.~\ref{cluster12vs1n2}, within the circled region, we see similar effects of nonlinearity when comparing Cluster-12 with Cluster-$3\&4\&5$, the union of the 3rd, 4th, and 5th bins from the left-hand side of the histogram in Fig.~\ref{CCSHCtree}. It is worth noting that the degree of mixing is slightly expanded compared to that observed in panel (A). Nonetheless, both heatmaps together echo the fact that the nonlinearity of the CCS system's dynamics is rather widespread.

\begin{figure}[ht!]
 \centering
\includegraphics[width=1.0\textwidth]{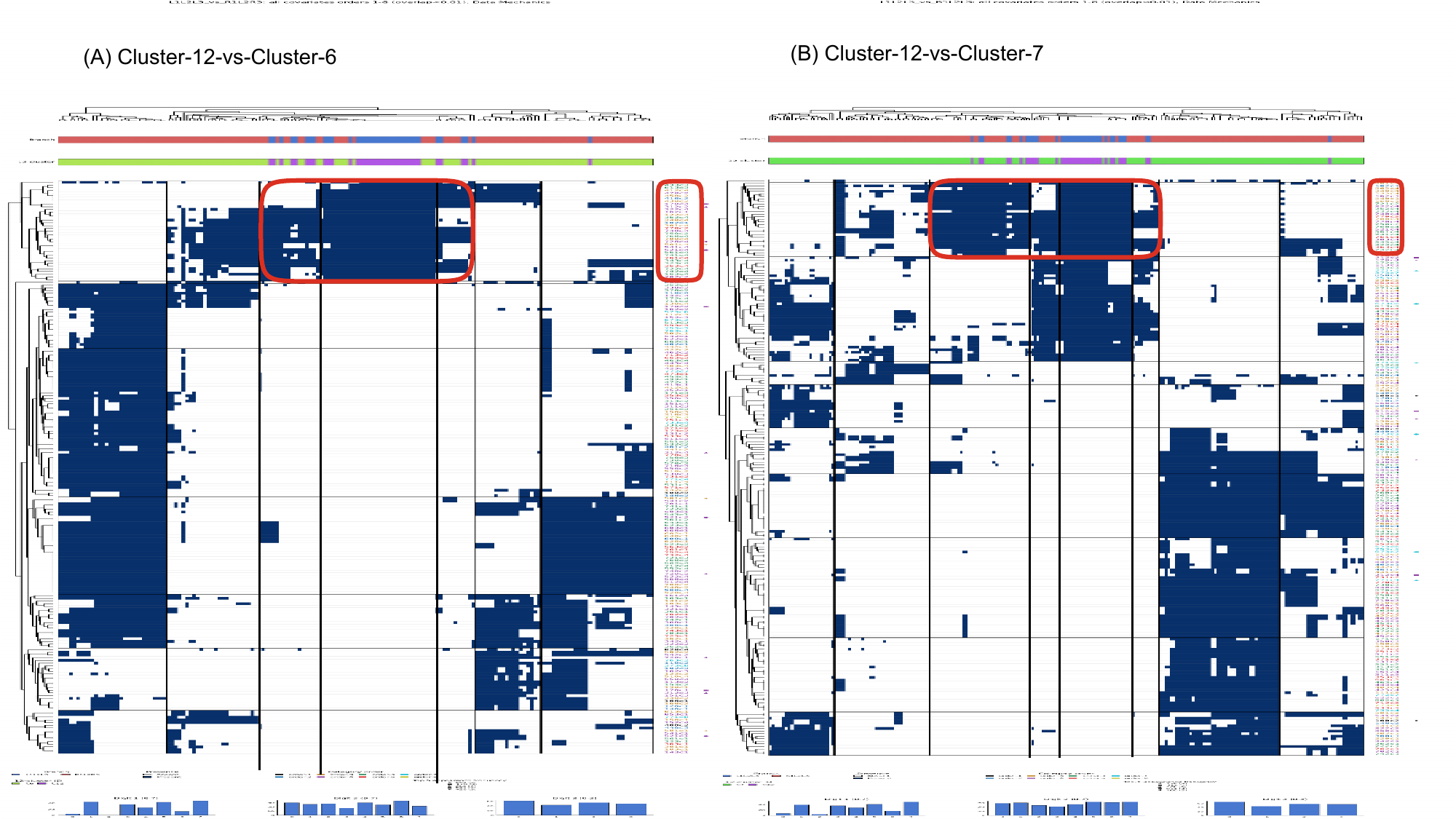}
 \caption{Two heatmaps exploring the potential for randomness or nonlinearity: (A) L1L2L3-vs-R1L2R3; (B) L1L2L3-vs-R1L2L3. }
 \label{cluster12vs6}
 \end{figure}

\begin{figure}[ht!]
 \centering
\includegraphics[width=1.0\textwidth]{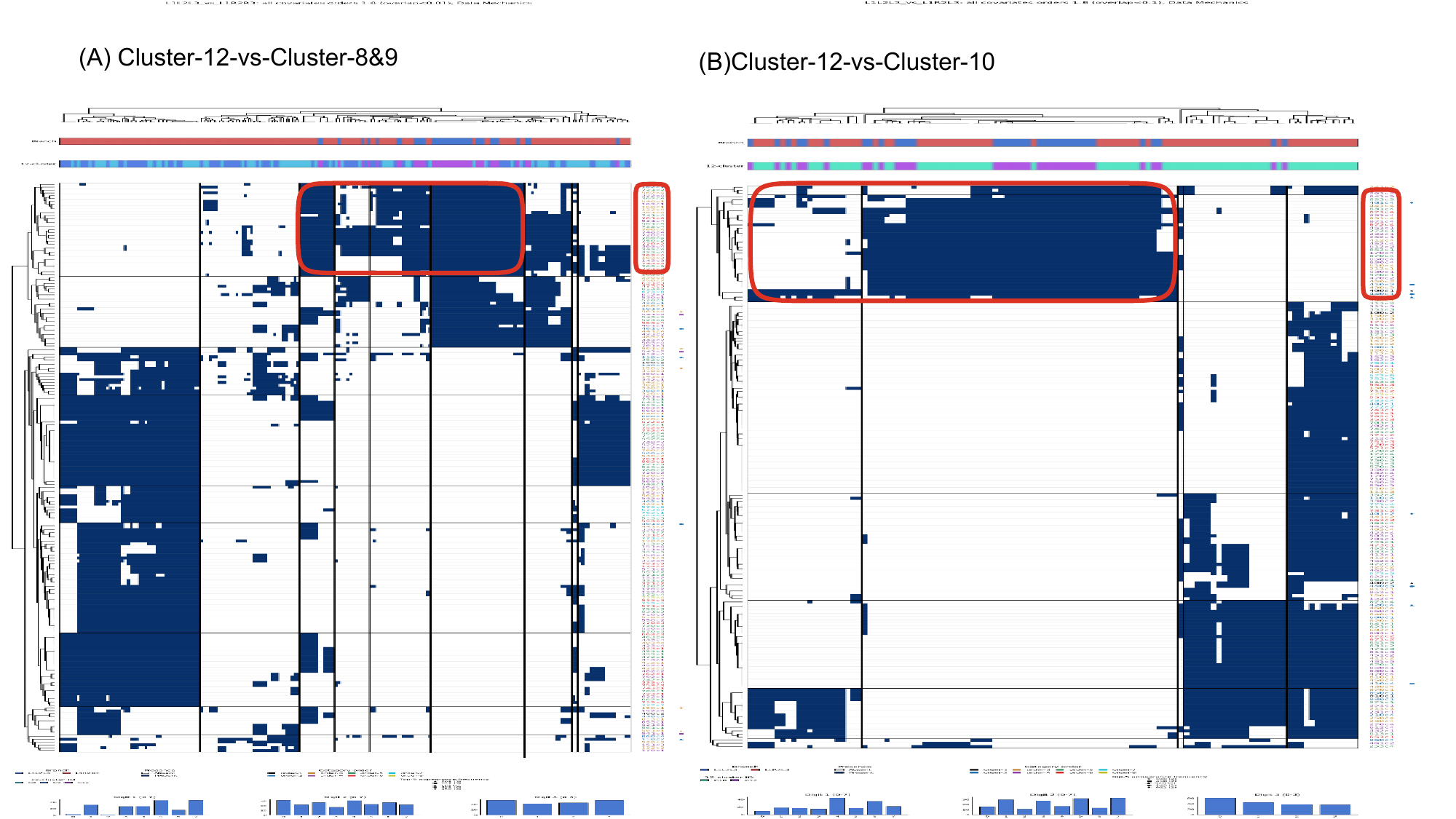}
 \caption{Two heatmaps exploring the potential for randomness or nonlinearity: (A) L1L2L3-vs-L1R2R3; (B) L1L2L3-vs-L1R2L3. }
 \label{cluster12vs8n9}
 \end{figure}

Similarly, nonlinearity appears to remain the primary factor driving the mixing patterns observed in the two panels of Fig.~\ref{cluster12vs6}, comparing Cluster-12 with Cluster-6 in panel (A) and with Cluster-7 in panel (B). These two heatmaps' mixing patterns are relatively close to that in panel (B) of Fig.~\ref{cluster12vs1n2}, but show some additional expansion through the blue-color-encoded members of Cluster-12. These observations suggest that the effects of randomness might begin to play a role here. A similarly slight expansion of the mixing pattern is seen when comparing Cluster-12 with Cluster-$8\&9$, in panel (A) of Fig.~\ref{cluster12vs8n9}. However, we can likewise conclude that nonlinearity remains the major factor for this mixing pattern, with randomness playing only a minor role.

Some visible expansion of the mixing pattern is seen in panel (B), when comparing Cluster-12 with Cluster-10. This mixing pattern is drastically different from that observed in panel (A). We project that randomness plays a key role, if not the primary role, in driving these mixing patterns. This evolving mixing pattern becomes logical and evident as clusters approach Cluster-12.

This evolving mixing pattern converges into a large-scale phenomenon of nonlinearity and randomness, as seen in panel (A) of Fig.~\ref{splitL1L2nL1R2}, under the setting of comparing Cluster-12 (L1L2L3) and Cluster-11 (L1L2R3). This evolution can be phrased as follows: the CCS-system dynamic governing Cluster-12 can mix with the dynamics governing other clusters, due to the interplay of randomness and nonlinearity. The degree of mixing is determined by how far a comparison cluster is from Cluster-12: the closer they are, the higher the degree of mixing due to randomness. Similar evolutions are observed in comparisons centered on clusters other than Cluster-12 (L1L2L3). Such evolutionary mixing patterns play an essential role in the DoE developed in the next section.

\section{Design of Experiment}
In this section, we discuss Experimental Design, or Design of Experiment (DoE), for any designated class belonging to the collection of ending nodes of the Taxonomic Hierarchy shown in Fig.~\ref{CCSTH}. Based on the spectrum of evolving mixing patterns across all pairwise comparisons discussed in the previous section, the effects of randomness and nonlinearity pertaining to a designated class dominate its DoE protocol. Since these effects are finite-sample- and targeted-class-specific, with no functional or distributional description, a universal DoE for all classes is not possible. As shown below, a DoE protocol for achieving a designated class within the Taxonomic Hierarchy is constructed based on a series of SDA heatmaps, following the principle below.\\

\paragraph{DoE principle:} {\bf Eliminate all sources of nonlinearity before handling randomness, in order to achieve the goal with reliability.}

This principle is needed because there is no idiosyncratic description for any designated class; any description of the targeted class is seen through comparisons. Therefore, this DoE protocol must be built along the following procedure: ``Abide by all constraints, inclusive and exclusive, to shape a covariate range leading to the designated class.'' Second, a reliable DoE protocol must include a built-in mechanism to avoid any sensitivity caused by nonlinearity, such as that demonstrated in panel (A) of Fig.~\ref{cluster12vs1n2}.

Following this principle, we first discuss the top-ranked class of CCS values in the Taxonomic Hierarchy. This DoE protocol is general-purpose: DoE for any designated class can follow the same protocol, with only slight modifications needed.

\subsection{DoE for the top-ranked CCS class}
There are 36 CCS measurements within cluster-12: 10 in class L1L2L3L4 (CCS-12L) and 26 in class L1L2L3R4 (CCS-12R), as shown in the Taxonomic Hierarchy in Fig.~\ref{CCSTH}. The CCS values in class CCS-12R are less than the CCS values in class CCS-12L, so CCS-12L is the top-ranked class. Our DoE quest is: How can we find a range of the 8 covariate features that gives rise to top-ranked CCS values with a calculated reliability?

It is worth reiterating that class CCS-12L does not have an explicit descriptive region or range within the 8-dimensional covariate feature space, since its 10 members are just 10 discrete points in that space. On the other hand, DoE for CCS-12L must be an explicit, descriptive region defined within the 8-dimensional space. We therefore make use of explicit, descriptive regions for CCS-12L when comparing it against various classes within the Taxonomic Hierarchy. One key advantage of this indirect description of CCS-12L is that it lets us identify which of its members are subject to sensitivity from nonlinearity. It is advantageous to exclude regions centered on such members when deriving the DoE range for CCS-12L.

\begin{figure}[ht!]
 \centering
\includegraphics[width=1.0\textwidth]{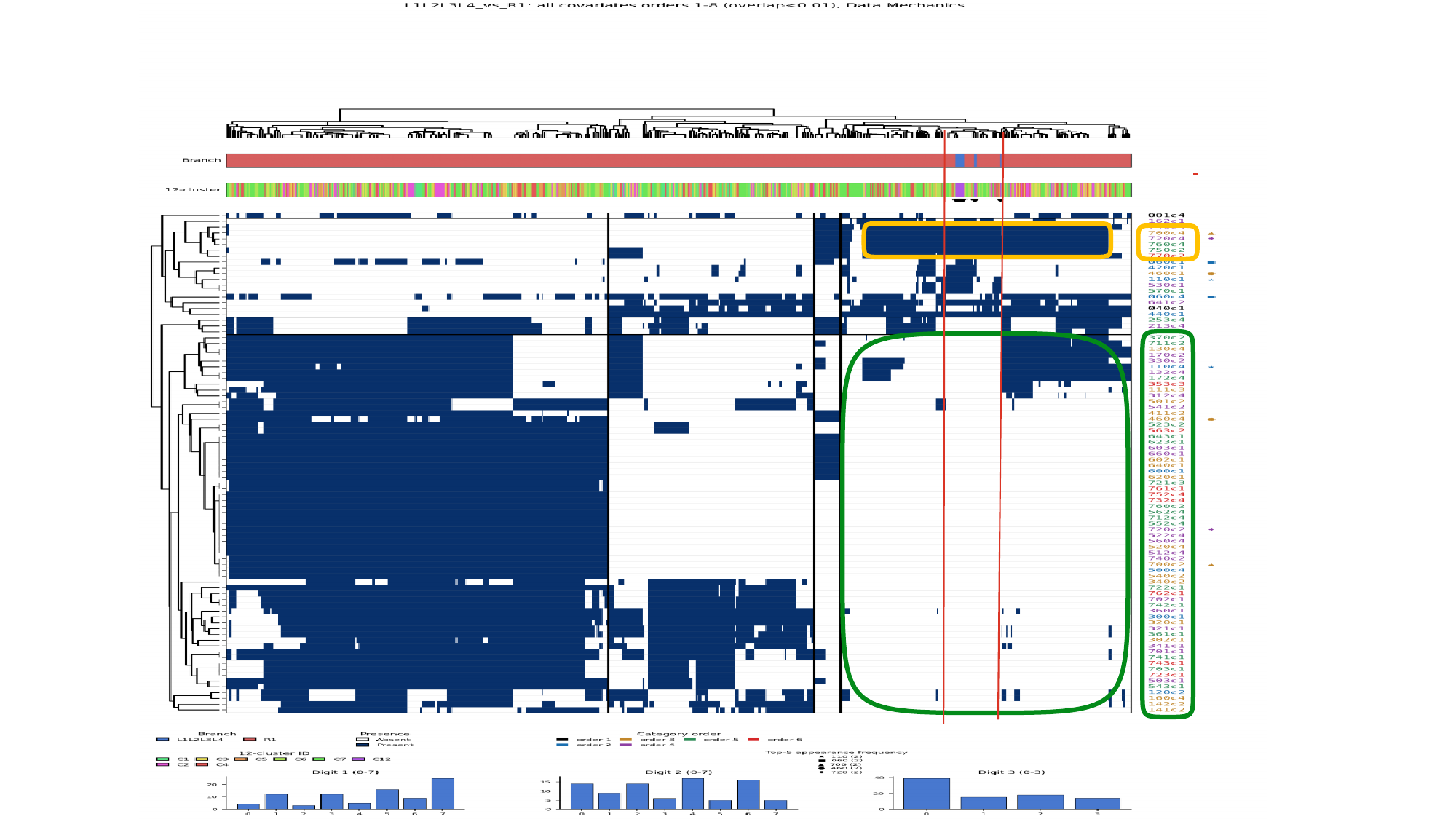}
 \caption{Design of Experiment for reduced ${\cal R}(L1L2L3L4)$ via the heatmap of L1L2L3L4-vs-R1.}
 \label{DoEL1L2L3L4vsR1}
 \end{figure}

Based on the Taxonomic Hierarchy shown in Fig.~\ref{CCSTH}, we propose the following sequential, four-step protocol.
\paragraph{CCS-12L against R1.}
We want a description of L1L2L3L4 (CCS-12L) against R1, so that we can exclude any potential sensitivity due to nonlinearity with respect to any classes under R1. An alternative approach for this step is to compare against all 8 classes in R1 individually. Based on the heatmap in Fig.~\ref{DoEL1L2L3L4vsR1}, we do not see high potential sensitivity due to nonlinearity from R1, since the members of L1L2L3L4 aggregate within a relatively small branch, mingling with only some members of R1.

To describe the DoE region for this small branch, we take the region of covariate space constrained by all major feature-set-categories within the orange circle, intersected with the complement of all major feature-set-categories within the green circle. See the Supporting Information (SI) for the full names of all major feature-sets and their category-IDs contained in the orange and green circles. Here, the orange circle consists of 6 major feature-set-categories with coded names and category-IDs: $\{740c4, 700c4, 720c4, 760c4, 750c2, 770c2\}$. Their full names are given below.
\begin{description}
\item[1]740c4=[Cement][Age][Fly Ash][Superplasticizer]c4,
\item[2]700c4=[Cement][Age][Fly Ash]c4,
\item[3]720c4=[Cement][Age][Fly Ash][Water]c4,
\item[4]760c4=[Cement][Age][Fly Ash][Water][Superplasticizer]c4,
\item[5]750c2=[Cement][Age][Fly Ash][Blast Furnace Slag][Superplasticizer]c2,
\item[6]770c2=[Cement][Age][Fly Ash][Blast Furnace Slag][Water][Superplasticizer]c2.
\end{description}
For instance, 740c4 represents an interacting effect of order 4. It is identified as cluster-c4 of the HC-tree built on the 4-dimensional data subset -- the union of L1L2L3L4 and R1 -- with respect to the four features $\{Cement, Age, Fly Ash, Superplasticizer\}$. As such, 740c4 serves as one inclusive constraint on the DoE range of L1L2L3L4. That is, this DoE range, as a covariate subspace, must satisfy 6 inclusive constraints indicated by the orange circle, and simultaneously 65 exclusive constraints indicated by the green circle; see SI.

\begin{figure}[ht!]
 \centering
\includegraphics[width=1.0\textwidth]{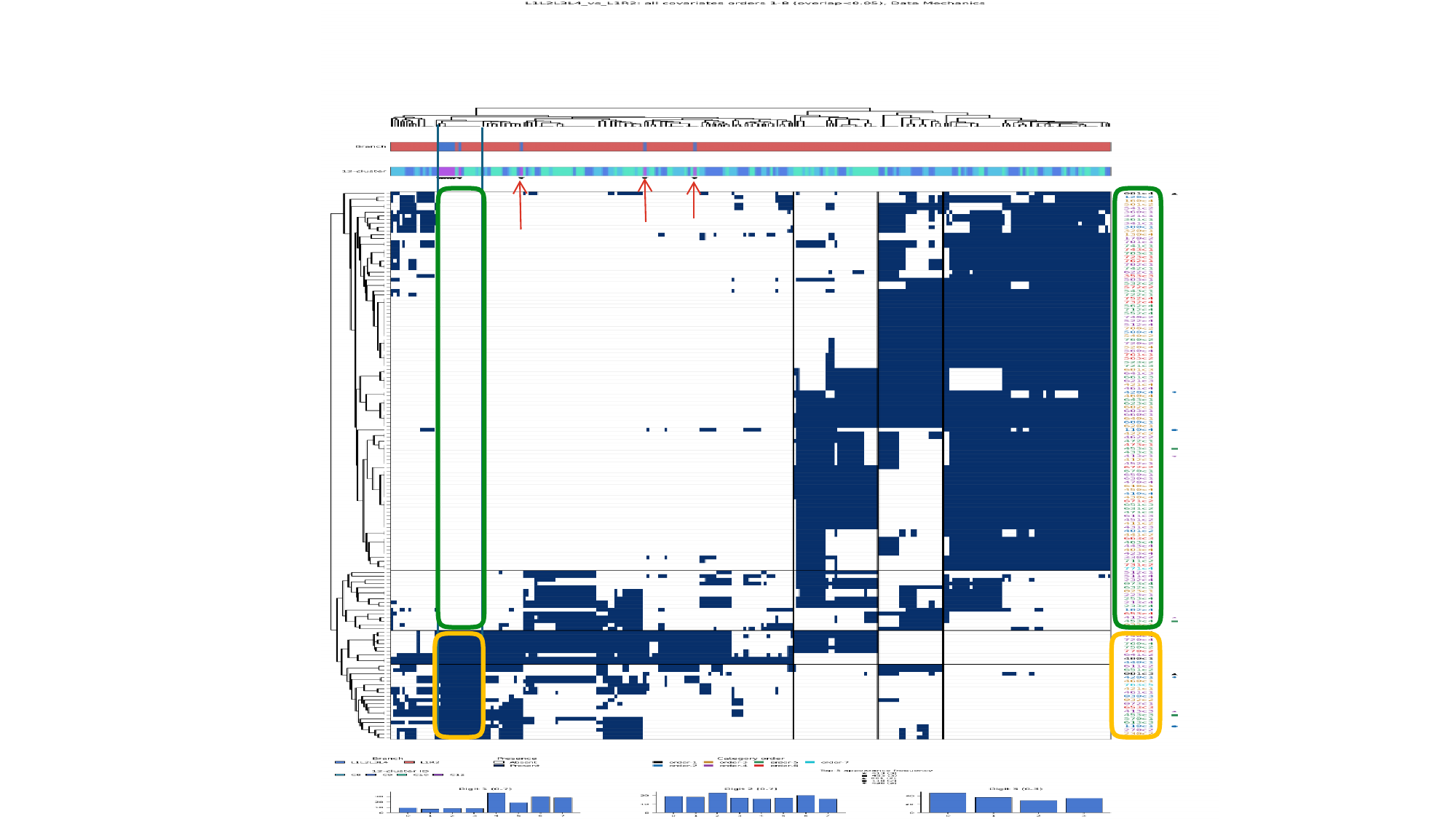}
 \caption{Design of Experiment for reduced ${\cal R}(L1L2L3L4)$ via the heatmap of L1L2L3L4-vs-L1R2.}
 \label{DoEL1L2L3L4vsL1R2}
 \end{figure}

\paragraph{CCS-12L against L1R2.}
Having found no potential R1-oriented sensitivity within CCS-12L, we proceed to derive a description of L1L2L3L4 (CCS-12L) against L1R2, which contains 2 classes. Based on the heatmap in Fig.~\ref{DoEL1L2L3L4vsL1R2}, we see that 3 members of CCS-12L (marked with red arrows) appear to result from nonlinearity, since they have rather distinct binary column-vectors from the remaining 7 members. We therefore exclude these three from CCS-12L, due to their potential sensitivity to nonlinearity with respect to classes under L1R2.

The description of the ``reduced CCS-12L'' is well characterized by the block-chain: a green block of 0's followed by an orange block of 1's. We again take the region of covariate space constrained by all major feature-set-categories within the orange circle, intersected with the complement of all major feature-set-categories within the green circle.

This DoE range, as a covariate subspace, must further satisfy 39 inclusive constraints indicated by the orange circle, and simultaneously 117 exclusive constraints indicated by the green circle; see SI.

\begin{figure}[ht!]
 \centering
\includegraphics[width=1.0\textwidth]{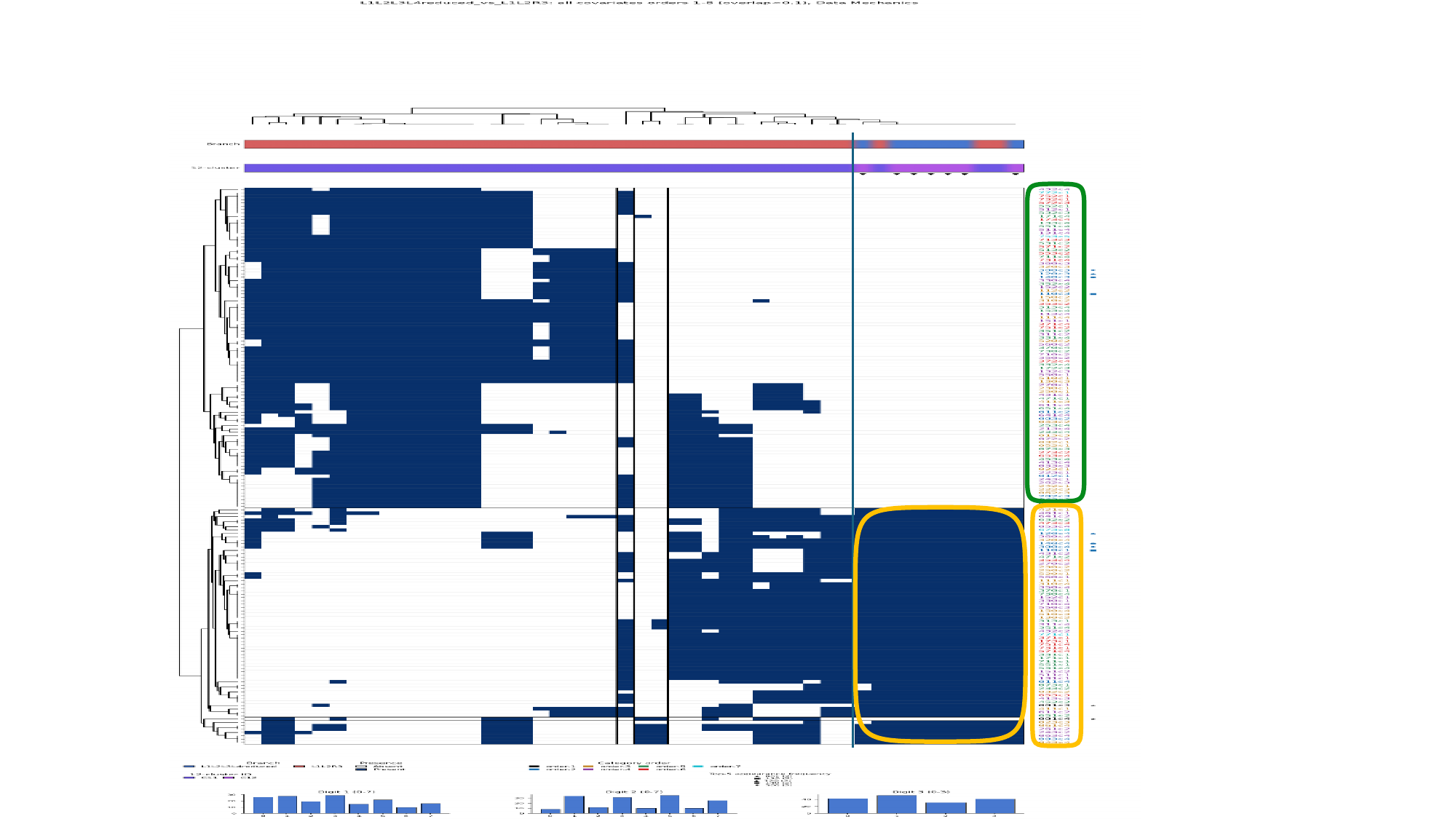}
 \caption{Design of Experiment for reduced ${\cal R}(L1L2L3L4)$ via the heatmap of L1L2L3L4-vs-L1L2R3.}
 \label{DoEL1L2L3L4vsL1L2R3}
 \end{figure}

\paragraph{Reduced CCS-12L against L1L2R3.}
Having eliminated three members of L1L2L3L4 due to their potential L1R2-sensitivity, we proceed to derive a description of the ``reduced L1L2L3L4'' against L1L2R3, which contains two classes: CCS-11L and CCS-11R. Based on the heatmap in Fig.~\ref{DoEL1L2L3L4vsL1L2R3}, all 7 remaining members share a common column-vector. We therefore conclude that there is no potential sensitivity to nonlinearity with respect to classes under L1L2R3. Again, we take the region of covariate space constrained by all major feature-set-categories within the orange circle, intersected with the complement of all major feature-set-categories within the green circle.

This DoE range, as a covariate subspace, must further satisfy 68 inclusive constraints indicated by the orange circle, and simultaneously 95 exclusive constraints indicated by the green circle; see SI.

\begin{figure}[ht!]
 \centering
\includegraphics[width=1.0\textwidth]{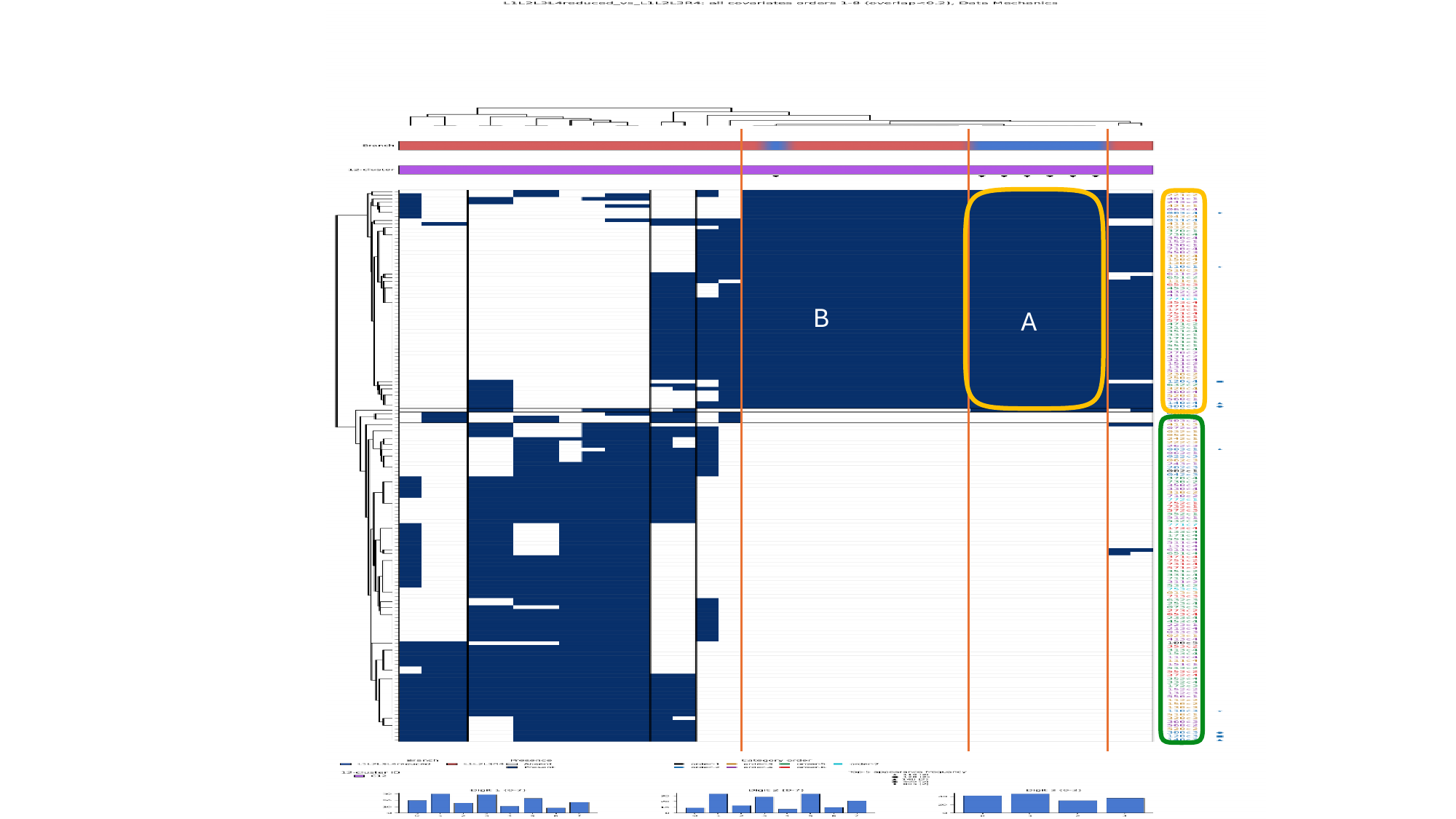}
 \caption{Design of Experiment for reduced ${\cal R}(L1L2L3L4)$ via the heatmap of L1L2L3L4-vs-L1L2L3R4.}
 \label{DoEL1L2L3L4vsL1L2L3R4}
 \end{figure}

\paragraph{Reduced CCS-12L against L1L2L3R4.}
Finally, we want a description of the ``reduced L1L2L3L4'' against L1L2L3R4 (CCS-12R). Based on the heatmap in Fig.~\ref{DoEL1L2L3L4vsL1L2L3R4}, we find that one member of the reduced L1L2L3L4 has a slightly distinct column-vector, mingling with a group of members of CCS-12R, as shown in block B. We take this member to be CCS-12-sensitive, and therefore exclude it as well from the reduced L1L2L3L4. The remaining 6 members, termed the ``2nd-reduced L1L2L3L4,'' are shown as exclusively occupying block A -- that is, characterized by a perfect block-chain. We again take the region of covariate space constrained by all major feature-set-categories within the orange circle, intersected with the complement of all major feature-set-categories within the green circle.

This DoE range, as a covariate subspace, must further satisfy 61 inclusive constraints indicated by the orange circle, and simultaneously 89 exclusive constraints indicated by the green circle; see SI.

In conclusion, the DoE for top-ranked CCS values via class L1L2L3L4 is delivered by the range resulting from intersecting the four regions above, excluding potential nonlinearity sensitivity with respect to four groups of classes. The final description of the 2nd-reduced L1L2L3L4 guarantees that this final DoE range will deliver only top-ranked CCS values, with $100\%$ reliability.

\subsection{DoE in general}
At the end of this section, the DoE for the top-ranked CCS-value class can be readily extended to any other class in the Taxonomic Hierarchy shown in Fig.~\ref{CCSTH}. For instance, the top-ranked CCS-value cluster shares its parent branch with the 2nd-ranked CCS-value cluster. As such, DoE aiming at the 2nd-ranked class is likewise defined by: How can we find a range of the 8-dimensional covariate features that specifically gives rise to CCS values falling into the 2nd-ranked class, with reliability?

Since both the top- and 2nd-ranked CCS-value classes share a common parent branch -- though their locality-specific randomness is distinct -- the DoE for the 2nd-ranked CCS-value class would follow the same protocol for exploring nonlinearity sensitivity laid out in this section. This is our CT-based DoE aimed at the class of 2nd-ranked CCS values.

More generally, the protocol for DoE aimed at different classes proceeds correspondingly: first finding potential nonlinearity in the farthest class, then the 2nd-farthest class, and so on. We can therefore conclude that any targeted class within the Taxonomic Hierarchy follows a similar protocol for its DoE.

\section{Conclusions}
Based on the pilot study of Concrete Compressive Strength (CCS), we constructed a Taxonomic Hierarchy for our development of Design of Experiment (DoE). It is a brand-new theme of working with real-world complex systems. Its fundamental role in DoE is seen through its representation of the underlying dynamics of the CCS system, via a collection of randomness-bearing classes and numerous evolutions of nonlinearity-bearing mixing patterns, revealed through SDA-based heatmaps across all class-vs-class comparisons. Specifically, the locality-specific randomness embraced by each class is defined as coherence of homogeneity between the response and covariate features under a single-class setting, while the multiscale nonlinearity is revealed through incoherence of homogeneity between the response and covariate features under settings involving multiple classes. It is essential to recognize that a class stands for locality-specific homogeneity -- also called pure randomness at the finest scale -- pertaining to the finite sample of the data set.

It is worth reiterating that a DoE quest is scientifically meaningful only when its designated class is of this finite-sample nature. This statement has a counterpart in classification: a classification task is scientifically meaningful only within the data-driven, computed framework of the Taxonomic Hierarchy. Only scientifically meaningful classes do make the classification task scientific. Further, all system-specific nonlinearity and randomness embraced within Taxonomic Hierarchy are visible, readable and explainable, so is DoE. This brings out an entirely distinct foundation for DoE.

This is a brand-new, crucial point for science, since scientists are now free of man-made functional structures and distributional randomness. Furthermore, it is necessary to point out one significant fact that classic Experimental Design built with top-down modeling via structured functional forms coupled with prescribed randomness can not accommodate the ``unexpected'' kinds of nonlinearity observed and discussed here within CCS-system.

Throughout this paper, each comparison of branch-vs-branch, class-vs-class, or even class-split against class-split, is translated into one Re-Co dynamic, upon which SDA is applied and a heatmap is constructed. Further, when such a Re-Co-dynamic-specific heatmap is sustained by a block-structure, each of its vertical block-chains characterizes study-subjects belonging to its corresponding cluster on the column-axis, while its horizontal block-chains manifest mechanistic dependence among selected major feature-categories of various orders belonging to its corresponding cluster on the row-axis. Such characterization and manifestation reveal the Re-Co dynamic specific aspect of system dynamics.

This understanding reflects our resolution to the fundamental question: How does a Taxonomic Hierarchy, as a heterogeneity-vs-randomness map embraced within a pilot study, determine the DoE? This resolution also addresses the equivalent question: How does Computational Taxonomy (CT) give rise to the scientific basis for DoE? As such, the reasoning behind our CT-based DoE theme can be seen clearly.

This brand-new, scientific DoE works for all data types, quantitative and qualitative. It also works for response variables in functional, time-series, and image formats, because CT is a universal paradigm for data analysis. With this wide applicability, we hope this CT-based DoE will help scientists advance their sciences without being limited by the technicalities -- such as various forms of optimality -- imposed by classic Experimental Design.

It is worth mentioning that Taxonomic Hierarchy as our DoE basis is seemingly related to classic recursive partitioning methods \cite{morgan,sethi}. These methods have become very popular statistical tools used by scientists in many fields. The most well-known is CART \cite{breiman}, see a thorough review \cite{loh}. While our Taxonomic Hierarchy is built recursively based on heterogeneity-vs-homogeneity map, CART and all related partition-wise models are built by recursively constructing a tree for the data space until all data within a given region exhibit ``homogeneous behavior'' (defined in terms of minimizing the residual sum of squares) \cite{cheungauelee}. This ``homogeneity'' is rather distinct from the idea of homogeneity in this paper, which is defined and confirmed via heatmap as a bipartite network between study subjects and associative information pieces.

More importantly, the concept of randomness found within a class is exhibited through the coherence between locality-specific response and covariate homogeneity. In contrast, CART splits the data by partitioning covariate features alone, recursively, and the homogeneity is decided based on predictive results.

Further, the nonlinearity in this paper is visualized by comparing different classes and inferring the sensitivity of system dynamics under study. Different classes bear different nonlinearity. This is not the case for nonlinearity in partition models, which is a global structure simply selected with respect to a certain loss function resulting from partitioning the space of data \cite{cheungauelee}. In summary, CART and related recursive partitioning methods are all predictive. Our fundamental concepts: randomness and nonlinearity, of DoE are aiming for better understanding the dynamics underlying the system of interest. Predictive inferences are basically byproducts as being emphasized here and in \cite{CTonIris,CTonpenguin}.

Lastly, we make a final remark on the CCS system's dynamics based on all the heatmaps presented in this paper, in particular through the blocks marked with various colored frames. From the SI, we see various blocks displaying the names and category-IDs of all major feature-set-categories. Only a few names, not including category-IDs, are repeated within each heatmap. This is strikingly distinct from the heatmaps presented in \cite{CTonIris} for Iris and \cite{CTonpenguin} for penguin data. Each of these two biological systems is characterized by a single group of major feature-sets that appears again and again, with different phases, within each heatmap -- apparently reflecting the force of natural selection. In contrast, there is no force of natural selection in the CCS system; all effects appear in a dispersed fashion. This striking distinction suggests that the CCS system has not been shaped or dominated by any single mechanism of civil engineering, which may itself be a defining characteristic of the CCS system.

%\pagebreak
%\reftitle{References}
\bibliographystyle{unsrt}

\end{document}